\documentclass[aps,eqsecnum,preprint,floats,epsf,epsfig,nofootinbib,letter]{revtex4}
\usepackage{epsfig}

\begin{document}
\def\be{\begin{eqnarray}}
\def\en{\end{eqnarray}}
\def\non{\nonumber}
\def\la{\langle}
\def\ra{\rangle}
\def\nc{N_c^{\rm eff}}
\def\vp{\varepsilon}
\def\drho{\bar\rho}
\def\deta{\bar\eta}
\def\CP{{\it CP}~}
\def\a{{\cal A}}
\def\B{{\cal B}}
\def\c{{\cal C}}
\def\d{{\cal D}}
\def\e{{\cal E}}
\def\p{{\cal P}}
\def\t{{\cal T}}
\def\up{\uparrow}
\def\dw{\downarrow}
\def\vma{{_{V-A}}}
\def\vpa{{_{V+A}}}
\def\smp{{_{S-P}}}
\def\spp{{_{S+P}}}
\def\J{{J/\psi}}
\def\ov{\overline}
\def\Lqcd{{\Lambda_{\rm QCD}}}
\def\pr{{Phys. Rev.}~}
\def\prl{{Phys. Rev. Lett.}~}
\def\pl{{Phys. Lett.}~}
\def\np{{Nucl. Phys.}~}
\def\zp{{Z. Phys.}~}
\def\lsim{ {\ \lower-1.2pt\vbox{\hbox{\rlap{$<$}\lower5pt\vbox{\hbox{$\sim$}
}}}\ } }
\def\gsim{ {\ \lower-1.2pt\vbox{\hbox{\rlap{$>$}\lower5pt\vbox{\hbox{$\sim$}
}}}\ } }

\font\el=cmbx10 scaled \magstep2{\obeylines\hfill March, 2013}

\vskip 1.5 cm

\centerline{\large\bf Revisiting charmless hadronic $B$ decays to
scalar mesons}

\bigskip
\centerline{\bf Hai-Yang Cheng$^{1}$, Chun-Khiang Chua$^{2}$,
Kwei-Chou Yang$^{2}$, Zhi-Qing Zhang$^{1,3}$}
\medskip
\centerline{$^1$ Institute of Physics, Academia Sinica}
\centerline{Taipei, Taiwan 115, Republic of China}
\medskip
\centerline{$^2$ Department of Physics, Chung Yuan Christian
University} \centerline{Chung-Li, Taiwan 320, Republic of China}
\medskip
\centerline{$^3$ Department of Physics, Henan University of Technology} \centerline{Zhengzhou, Henan 450052, P.R. China}
\bigskip
\bigskip
\bigskip
\bigskip
\bigskip
\centerline{\bf Abstract}
\bigskip
\small
Hadronic charmless $B$ decays to scalar mesons are studied within the framework of QCD factorization (QCDF).
Considering two different scenarios for scalar mesons above 1 GeV, we find that the data favor the scenario in which the scalars $a_0(1450)$ and $K_0^*(1430)$ are the lowest lying $q\bar q$ bound states. This in turn implies a preferred four-quark nature for light scalars below 1 GeV.
Assuming $K_0^*(1430)$ being a lowest lying $q\bar s$ state, we show that the data of $B\to K_0^*(1430)\eta^{(')}$ and $B\to K_0^*(1430)(\rho,\omega,\phi)$ can be accommodated in QCDF without introducing power corrections induced from penguin annihilation,   while the predicted $B^-\to \ov K_0^{*0}(1430)\pi^-$ and $\ov B^0\to K_0^{*-}(1430)\pi^+$ are too small compared to experiment.
In principle, the data of $K_0^*(1430)\pi$ modes can be explained if penguin-annihilation induced power corrections are taken into account. However, this will destroy the agreement between theory and experiment for $B\to K_0^*(1430)(\eta^{(')},\rho,\omega,\phi)$.
Contrary to the pseudoscalar meson sector where $B\to K\eta'$ has the largest rate in 2-body decays of the $B$ meson, we show that $\B(B\to K_0^*\eta')<\B(B\to K_0^*\eta)$.
The decay $\ov B^0\to a_0(980)^+K^-$ is found to have a rate much smaller than that of $\ov B^0\to a_0(980)^+\pi^-$ in QCDF, while it is the other way around in pQCD. Experimental measurements of these two modes will help discriminate between these two different approaches. Assuming 2-quark bound states for $f_0(980)$ and $f_0(500)$, the observed large rates of $f_0(980)K$ and $f_0(980)K^*$ modes can be explained in QCDF with the $f_0(980)\!-\!f_0(500)$ mixing angle $\theta$ in the vicinity of $20^\circ$. However, this does not necessarily imply that a 4-quark assignment for $f_0(980)$ is ruled out because of extra diagrams contributing to $B\to f_0(980)K^{(*)}$. Irrespective of the mixing angle $\theta$, the predicted branching fraction of $B^0\to f_0(980)\rho^0$ is far below the Belle measurement and this needs to be clarified in the future.

\pagebreak

\section{Introduction}

In the past few years there are some progresses in the study of charmless hadronic $B$ decays with scalar mesons in the final state both experimentally and theoretically. On the experimental side, measurements of $B$ decays to the scalar mesons such as $f_0(980),f_0(1370),f_0(1500),f_0(1710)$, $a_0(980),a_0(1450)$ and $K_0^*(1430)$ have been reported by BaBar and Belle; see Tables \ref{tab:exptBR} and \ref{tab:exptCP} for a summary of the experimental results. It is well known that the
identification of scalar mesons is difficult experimentally and
the underlying structure of scalar mesons is not well established
theoretically. The experimental measurements of
$B\to SP$ and $B\to SV$, where $S,P,V$ stand for scalar, vector and pseudoscalar mesons, respectively, will provide valuable information on the nature of the even-parity mesons.
On the theoretical side, hadronic $B$ decays to scalar mesons have been studied in the QCD-inspired approaches: QCD factorization (QCDF) \cite{CCY:SP,CCY:SV,CC:Keta,Li:K0stpi,Li:K0strho} and pQCD \cite{WangWei,Shen,Zhang:f0pi,Xiao:K0steta,Zhang:K0strho,Zhang:f0Kst,Zhang:a0pi,CSKim,Zhang:a01450rho,Zhang:a01450Kst}.

In this work, we would like to revisit the study of the 2-body charmless decays $B\to SP$ and $B\to SV$ within the framework of QCDF for the following reasons: (i) In \cite{CCY:SP} we have  missed some factorizable terms (more precisely, the $f_0$ and $a_0^0$ emission terms) in the expressions for the decay amplitudes of $B\to
f_0K,~a_0^0\pi,~a_0^0K$. (ii) Attention has not been paid to the relative sign difference of the vector decay constants between $a_0^-$ and $a_0^+$ and between $K_0^*$ and $\ov K_0^*$ or $K_0^{*-}$ and $K_0^{*+}$ in our previous study. (iii) There were some errors in our previous computer code which may significantly affect some of the calculations done before.  (iv) Progress has been made in the past in the study of $B\to S$ transition form factors in various approaches  \cite{Aliev,Lu:2008,Wang,Sun,Han,Lu:2009,Xiao:K0steta}.
(v) Experimental data for some of $B\to SV$ decays such as $K_0^*(1430)\phi, K_0^*(1430)\rho$ and $K_0^*(1430)\omega$ are now available.
(vi) It is known that in order to account for the penguin-dominated $B\to PP,VP,VV$ decay modes within the framework of QCDF, it is necessary to include power corrections due to penguin annihilation \cite{BBNS,BN}. In the present work, we wish to examine if the same effect holds in the scalar meson sector; that is, if the penguin-annihilation induced power corrections are also needed to explain the penguin dominated $B\to SP$ and $B\to SV$ decays.

\begin{table}[!]
\caption{Experimental branching fractions (in units of
$10^{-6}$) of $B$ decays to scalar mesons \cite{HFAG}.
} \label{tab:exptBR}
\scriptsize{
\begin{ruledtabular}
\begin{tabular}{l c c c}
Mode & BaBar & Belle & Average  \\
\hline
 $\B(B^+\to f_0(980)K^+)\B(f_0(980)\to\pi^+\pi^-)$ & $10.3\pm0.5^{+2.0}_{-1.4}$ & $8.8\pm0.8^{+0.9}_{-1.8}$ &  $9.4^{+0.9}_{-1.0}$ \\
 $\B(B^0\to f_0(980)K^0)\B(f_0(980)\to\pi^+\pi^-)$ & $6.9\pm0.8\pm0.6$ & $7.6\pm1.7^{+0.9}_{-1.3}$ & $7.0\pm0.9$  \\
 $\B(B^+\to f_0(980)K^+)\B(f_0(980)\to\pi^0\pi^0)$ & $2.8\pm0.6\pm0.5$ &
 &  $2.8\pm0.8$ \\
 $\B(B^+\to f_0(980)K^+)\B(f_0(980)\to K^+K^-)$ & $9.4\pm1.6\pm2.8$ &
 $<2.9$  \\
 $\B(B^0\to f_0(980)K^0)\B(f_0(980)\to K^+K^-)$ &
 $7.0^{+3.5}_{-3.0}$ & & $7.0^{+3.5}_{-3.0}$   \\
 $\B(B^+\to f_0(980)\pi^+)\B(f_0(980)\to\pi^+\pi^-)$ &
 $<1.5$ & & $<1.5$  \\
 $\B(B^0\to f_0(980)\eta')\B(f_0(980)\to\pi^+\pi^-)$ & $<0.9$ & &  $<0.9$ \\
 $\B(B^0\to f_0(980)\eta)\B(f_0(980)\to\pi^+\pi^-)$ &
 $<0.4$ & & $<0.4$ \\
 $\B(B^+\to a_0^0(980)K^+)\B(a_0(980)^0\to\eta\pi^0)$ & $<2.5$ & & $<2.5$  \\
 $\B(B^0\to a_0^-(980)K^+)\B(a_0(980)^-\to\eta\pi^-)$ & $<1.9$ & & $<1.9$ \\
 $\B(B^+\to a_0^+(980)K^0)\B(a_0(980)^+\to\eta\pi^+)$ & $<3.9$ & & $<3.9$  \\
 $\B(B^0\to a_0^0(980)K^0)\B(a_0(980)^0\to\eta\pi^0)$ & $<7.8$  & & $<7.8$ \\
 $\B(B^+\to a_0^0(980)\pi^+)\B(a_0(980)^0\to\eta\pi^0)$ & $<5.8$ & & $<5.8$ \\
 $\B(B^+\to a_0^+(980)\pi^0)\B(a_0(980)^0\to\eta\pi^0)$ & $<1.4$ & & $<1.4$ \\
 $\B(B^0\to a_0^\mp(980)\pi^\pm)\B(a_0(980)^\mp\to\eta\pi^\mp)$ & $<3.1$ & & $<3.1$ \\
 $\B(B^0\to a_0^\mp(1450)\pi^\pm)\B(a_0(1450)^\mp\to\eta\pi^\mp)$ & $<2.3$ & & $<2.3$ \\
 $\B(B^0\to a_0^-(1450)K^+)\B(a_0(1450)^-\to \eta \pi^-)$ & $<3.1$ & & $<3.1$  \\
 $\B(B^+\to f_0(1370)K^+)\B(f_0(1370)\to\pi^+\pi^-)$ & $<10.7$ & & $<10.7$ \\
 $\B(B^+\to f_0(1370)\pi^+)\B(f_0(1370)\to\pi^+\pi^-)$ & $2.9\pm0.5\pm0.5^{+0.7}_{-0.5}<4.0$  & & $<4.0$\\
 $\B(B^+\to f_0(1500)K^+)$ & $17\pm4\pm12$ & & $17.0\pm12.6$ \\
 $\B(B^0\to f_0(1500)K^0)$ & $13.3^{+5.8}_{-4.4}\pm3.2$ & & $13.3^{+6.6}_{-5.4}$ \\
 $\B(B^+\to f_0(1710)K^+)\B(f_0(1710)\to K^+K^-)$ & $1.12\pm0.25\pm0.50$ & $1.7\pm1.0$ & $1.26\pm0.49$ \\
 $\B(B^0\to f_0(1710)K^0)\B(f_0(1710)\to K^+K^-)$ & $4.4\pm0.7\pm0.5$ & & $4.4\pm0.9$  \\
 $\B(B^0\to f_0(1710)K_S)\B(f_0(1710)\to K_SK_S )$ & $0.50^{+0.46}_{-0.24}\pm0.11$ & & $0.50^{+0.47}_{-0.26}$ \\
 $\B(B^+\to K^{*0}_0(1430)\pi^+)$ & $32.0\pm1.2^{+10.8}_{-~6.0}$ & $51.6\pm1.7^{+7.0}_{-7.5}$ &   $45.1\pm6.3$ \\
 $\B(B^0\to K^{*+}_0(1430)\pi^-)$ &
 $29.9^{+2.3}_{-1.7}\pm3.6$ \footnotemark[1] & $49.7\pm3.8^{+6.8}_{-8.2}$ & $33.5^{+3.9}_{-3.8}$ \\
 $\B(B^0\to K^{*0}_0(1430)\pi^0)$ &
 $7.0\pm0.5\pm1.1$  \footnotemark[2] & & $7.0\pm1.2$  \\
 $\B(B^+\to \ov K_0^{*0}(1430)K^+)$ & $<2.2$ & & $<2.2$ \\
 $\B(B^+\to K_0^{*+}(1430)\eta')$ & $5.2\pm1.9\pm1.0$ & & $5.2\pm2.1$ \\
 $\B(B^0\to K_0^{*0}(1430)\eta')$ & $6.3\pm1.3\pm0.9$ & & $6.3\pm1.6$ \\
 $\B(B^+\to K_0^{*+}(1430)\eta)$ & $15.8\pm2.2\pm2.2$ & & $15.8\pm3.1$  \\
 $\B(B^0\to K_0^{*0}(1430)\eta)$ & $9.6\pm1.4\pm1.3$ & & $9.6\pm1.9$ \\
 \hline
 $\B(B^+\to f_0(980)K^{*+})\B(f_0(980)\to\pi^+\pi^-)$ & $4.2\pm0.6\pm0.3$ &
 &  $4.2\pm0.7$ \\
 $\B(B^0\to f_0(980)K^{*0})\B(f_0(980)\to\pi^+\pi^-)$ & $5.7\pm0.6\pm0.3$ & $<2.2$  &  $5.7\pm0.7$ \\
 $\B(B^+\to f_0(980)\rho^+)\B(f_0(980)\to\pi^+\pi^-)$ &
 $<2.0$ & & $<2.0$  \\
 $\B(B^0\to f_0(980)\rho^0)\B(f_0(980)\to\pi^+\pi^-)$ &
 $<0.40$ & $0.87\pm0.27\pm0.15$ \footnotemark[3] & $0.87\pm0.31$  \\
 $\B(B^0\to f_0(980)\omega)\B(f_0(980)\to\pi^+\pi^-)$ &
 $<1.5$ & & $<1.5$ \\
 $\B(B^0\to f_0(980)\phi)\B(f_0(980)\to\pi^+\pi^-)$ &
 $<0.38$ & & $<0.38$ \\
 $\B(B^+\to K_0^{*+}(1430)\omega)$ & $24.0\pm2.6\pm4.4$ & & $24.0\pm5.1$ \\
 $\B(B^0\to K_0^{*0}(1430)\omega)$ & $16.0\pm1.6\pm3.0$ & & $16.0\pm3.4$ \\
 $\B(B^+\to K_0^{*+}(1430)\phi)$ & $7.0\pm1.3\pm0.9$ & & $7.0\pm1.6$ \\
 $\B(B^0\to K_0^{*0}(1430)\phi)$ & $3.9\pm0.5\pm0.6$ & & $3.9\pm0.8$ \\
 $\B(B^0\to K_0^{*0}(1430)\ov K^{*0})$ & & $<3.3$  & $<3.3$ \\
 $\B(B^0\to K_0^{*0}(1430)K^{*0})$ & $<1.7$ & & $<1.7$ \\
 $\B(B^0\to K_0^{*+}(1430)\rho^-)$ & $28\pm10\pm5\pm3$ & & $28.0\pm11.6$ \\
 $\B(B^0\to K_0^{*0}(1430)\rho^0)$ & $27\pm4\pm2\pm3$ & & $27.0\pm5.4$ \\
 \hline
 $\B(B^0\to f_0(980)f_0(980))\B^2(f_0(980)\to \pi^+\pi^-)$ & $<0.19$ & $<0.1$ & $<0.1$ \\
 $\B(B^0\to f_0(980)K_0^{*0}(1430))\B(f_0(980)\to\pi\pi)$ & $2.7\pm0.7\pm0.5\pm0.3$ & & $2.7\pm0.9$ \\
 $\B(B^0\to K_0^{*0}(1430)K_0^{*0}(1430))$ & & $<4.7$ & $<4.7$ \\
 $\B(B^0\to K_0^{*0}(1430)\ov K_0^{*0}(1430))$ & & $<8.4$ & $<8.4$ \\
\end{tabular}
\end{ruledtabular}
\footnotetext[1]{There is a new measurement of $(27.8\pm2.5\pm3.3)\times 10^{-6}$ extracted from $B^0\to K^+\pi^-\pi^0$ by BaBar \cite{BABAR:Kppimpi0}.}
\footnotetext[2]{See \cite{BABAR:Kppimpi0}.}
\footnotetext[3]{See \cite{Belle:f0rho0}.}
}
\end{table}

\begin{table}[!]
\caption{Experimental {\it CP} asymmetries (in units of \%) of $B$ decays to final states containing scalar mesons \cite{HFAG}.
} \label{tab:exptCP}
\begin{ruledtabular}
\begin{tabular}{l c c c}
Mode & BaBar & Belle & Average  \\
\hline
 $B^+\to f_0(980)K^+$ & $-10.6\pm5.0^{+3.6}_{-1.5}$ \footnotemark[1] & $-7.7\pm6.5^{+4.6}_{-2.6}$ &  $-9.5^{+4.9}_{-4.2}$ \\
 $B^0\to f_0(980)K^0$ & $-28\pm24\pm9$ \footnotemark[2] & $-30\pm29\pm11\pm9$ \footnotemark[2] &  $-29\pm20$ \\
  & $-8\pm19\pm3\pm4$ \footnotemark[3] & $-6\pm17\pm7\pm9$ \footnotemark[3] &  $-7\pm14$ \\
 $B^+\to f_0(1370)\pi^+$ & $72\pm15\pm14^{+7}_{-8}$ & & $72\pm22$ \\
 $B^+\to f_0(1500)K^+$ & $28\pm26^{+15}_{-14}$ & & $28^{+30}_{-29}$ \\
 $B^+\to K^{*0}_0(1430)\pi^+$ & $3.2\pm3.5^{+3.4}_{-2.8}$ & $7.6\pm3.8^{+2.8}_{-2.2}$ &   $5.5^{+3.4}_{-2.2}$ \\
 $B^0\to K^{*+}_0(1430)\pi^-$ &
 $7\pm14\pm1$ & &  $7\pm14$  \\
 $B^0\to K^{*0}_0(1430)\pi^0$ &
 $-15\pm10\pm4$ & & $-15\pm11$    \\
 $B^+\to K_0^{*+}(1430)\eta'$ & $6\pm20\pm2$ & & $6\pm20$ \\
 $B^0\to K_0^{*0}(1430)\eta'$ & $-19\pm17\pm2$ & & $-19\pm17$ \\
 $B^+\to K_0^{*+}(1430)\eta$ & $5\pm13\pm2$ & & $5\pm13$  \\
 $B^0\to K_0^{*0}(1430)\eta$ & $6\pm13\pm2$ & & $6\pm13$ \\
 \hline
 $B^+\to f_0(980)K^{*+}$ & $-15\pm12\pm3$  &
 &  $-15\pm12$ \\
 $B^0\to f_0(980)K^{*0}$ & $7\pm10\pm2$ &  &  $7\pm10$ \\
 $B^+\to K_0^{*+}(1430)\omega$ & $-10\pm9\pm2$ & & $-10\pm9$ \\
 $B^0\to K_0^{*0}(1430)\omega$ & $-7\pm9\pm2$ & & $-7\pm9$ \\
 $B^+\to K_0^{*+}(1430)\phi$ & $4\pm15\pm4$ & & $4\pm15$ \\
 $B^0\to K_0^{*0}(1430)\phi$ & $20\pm14\pm6$ & & $20\pm15$ \\
\end{tabular}
\end{ruledtabular}
\footnotetext[1]{This data is from the measurement of $B^+\to K^+\pi^+\pi^-$.
BaBar's measurements of $B^+\to K^+K^+K^-$ and $B^+\to K^+\pi^0\pi^0$ yield $A_{CP}(B^+\to f_0(980)K^+)=-(8\pm8\pm4)\%$ \cite{BaBar:KKK} and  $A_{CP}(B^+\to f_0(980)K^+)=(18\pm18\pm4)\%$ \cite{BaBar:Kmpi0pi0}, respectively.}
\footnotetext[2]{From $B^0\to K^+K^-K_S$.}
\footnotetext[3]{From $B^0\to \pi^+\pi^-K_S$.}
\end{table}

This paper is organized as follows. We specify in Sec. 2 various input parameters for scalar mesons, such as decay constants, form factors and light-cone distribution amplitudes. The relevant decay amplitudes are briefly discussed in Sec. 3. Results and detailed discussions are presented in Sec. 4. Conclusions are given in Sec. 5. We lay out the explicit decay amplitudes of $B^0\to (f_0,a_0^0)(K,\pi)$ in Appendix A.

\begin{table}[tb]
\caption{The scalar decay constant $\bar f_{S}$ (in units of MeV) and Gegenbauer moments $B_1$, $B_3$ and  in scenario 1 (left)
and scenario 2 (right)  at the scale $\mu=1$ GeV
obtained using the QCD sum rule method \cite{CCY:SP}. Decay constants and  Gegenbauer moments for excited states in scenario 2 are not listed here.
} \label{tab:moments}
\begin{ruledtabular}
\begin{tabular}{l  r r r |  r r r}
 & $\bar f_S$~~ & $B_1$~~~~ & $B_3$~~~~~ &  $\bar f_S$~~ & $B_1$~~~~ & $B_3$~~~~~ \\ \hline
 $f_0(980)$ & $370\pm20$ & $-0.78\pm0.08$ & $0.02\pm0.07$ & \\
 $a_0(980)$ & $365\pm20$ & $-0.93\pm0.10$ & $0.14\pm0.08$ \\
 $K_0^*(800)$ & $340\pm20$ & $-0.92\pm0.11$ & $0.15\pm0.09$ & \\
 $f_0(1500)$ & $-255\pm30$ & $0.80\pm0.40$ & $-1.32\pm0.14$ & $490\pm50$ & $-0.48\pm0.11$ & $-0.37\pm0.20$ \\
 $a_0(1450)$ & $-280\pm30$ & $0.89\pm0.20$ & $-1.38\pm0.18$ & $460\pm50$ & $-0.58\pm0.12$ & $-0.49\pm0.15$ \\
 $K_0^*(1430)$ & $-300\pm30$ & $0.58\pm0.07$ & $-1.20\pm0.08$ & $445\pm50$ &
 $-0.57\pm0.13$ & $-0.42\pm0.22$ \\
\end{tabular}
\end{ruledtabular}
\end{table}

\begin{table}[t]
\caption{Form factors of $B\to a_0(1450),K_0^*(1430)$
transitions obtained in the covariant light-front model
\cite{CCH} in scenario 1 (upper entry) and scenario 2 (lower entry).
} \label{tab:FF}
\begin{ruledtabular}
\begin{tabular}{| l c c  c || l c c  c |}
~~~$F$~~~~~
    & $F(0)$~~~~~
    &$a$~~~~~
    & $b$~~~~~~
& ~~~ $F$~~~~~
    & $F(0)$~~~~~
    & $a$~~~~~
    & $b$~~~~~~
 \\
    \hline
$F^{Ba_0(1450)}_1$
    & $0.26$
    & 1.57
    & 0.70
 &$F^{Ba_0(1450)}_0$
    & 0.26
    & $0.55$
    & 0.03 \\
    & $0.21$
    & 1.66
    & 1.00
 &
    & 0.21
    & $0.73$
    & 0.09
    \\
$F^{BK^*_0(1430)}_1$
    & $0.26$
    & 1.52
    & 0.64
&$F^{BK^*_0(1430)}_0$
    & 0.26
    & 0.44
    & 0.05
    \\
    & $0.21$
    & 1.59
    & 0.91
&
    & 0.21
    & 0.59
    & 0.09
    \\
\end{tabular}
\end{ruledtabular}
\end{table}

\section{Physical properties of scalar mesons}

In order to study the hadronic charmless $B$ decays containing a scalar meson in the final state, it is necessary to specify the quark content of the scalar meson. For scalar mesons above 1 GeV we have explored in \cite{CCY:SP} two possible scenarios in the QCD sum rule method, depending on
whether the light scalars $K_0^*(800),~a_0(980)$ and $f_0(980)$ are
treated as the lowest lying $q\bar q$ states or four-quark
particles: (i) In scenario 1, we treat
$K_0^*(800), a_0(980), f_0(980)$ as the lowest lying $q\bar q$ states, and
$K_0^*(1430), a_0(1450),f_0(1500)$ as the corresponding first
excited states, respectively,  and (ii) we assume in scenario 2 that $K_0^*(1430),
a_0(1450), f_0(1500)$ are the lowest lying $q\bar q$ resonances and the
corresponding first excited states lie between $(2.0\sim
2.3)$~GeV. Scenario 2 corresponds to the case that light scalar
mesons are four-quark bound states, while all scalar mesons are
made of two quarks in scenario 1.  Phenomenological studies in \cite{CCY:SP,CCY:SV} imply that scenario 2 is preferable, which will be also reinforced in this work. Indeed, lattice calculations have confirmed that $a_0(1450)$ and $K_0^*(1430)$ are lowest-lying $P$-wave $q\bar q$ mesons \cite{Mathur}, and indicated that $f_0(500)$ (or $\sigma$) and $K_0^*(800)$ (or $\kappa$) are $S$-wave tetraquark mesonia \cite{Prelovsek}. \footnote{However, a recent lattice calculation \cite{Alexandrou:2012rm} leads to an opposite conclusion.}

\subsection{Decay constants and form factors}
Decay constants of scalar, pseudoscalar and vector mesons are defined as
 \be \label{eq:Sdecayc}
&& \la S(p)|\bar q_2\gamma_\mu q_1|0\ra=f_S p_\mu,
 \qquad \la S|\bar q_2q_1|0\ra=m_S\bar f_S, \non \\
&& \la P(p)|\bar q_2\gamma_\mu\gamma_5 q_1|0\ra=-if_Pp_\mu, \qquad
 \la V(p)|\bar q_2\gamma_\mu q_1|0\ra=f_Vm_V\vp_\mu^*, \non \\
&&  \la V(p,\vp^*)|\bar q_2\sigma_{\mu\nu}q_1|0\ra=f_V^\bot(p_\mu
 \vp_\nu^*-p_\nu\vp_\mu^*).
 \en
For scalar mesons, the vector decay constant $f_S$ and the
scale-dependent scalar decay constant $\bar f_S$ are related by
equations of motion
\be \label{eq:EOM}
 \mu_Sf_S=\bar f_S, \qquad\quad{\rm with}~~\mu_S={m_S\over
 m_2(\mu)-m_1(\mu)},
\en
where $m_{2}$ and $m_{1}$ are the running current quark masses and
$m_S$ is the scalar meson mass.  For the neutral scalar mesons
$f_0$, $a_0^0$ and $\sigma$, $f_S$ vanishes owing
to charge conjugation invariance or conservation of vector
current, but the quantity $\bar f_S=\mu_S f_S$ remains finite.
It is straightforward to show from Eq. (\ref{eq:Sdecayc}) that the decay constants of the scalar meson and its antiparticle  are related by
\be
 \bar f_{\bar S}=\bar f_S, \qquad f_{\bar S}=-f_S.
\en
Indeed, from Eq. (\ref{eq:EOM}) we have, for example,
\be \label{eq:a0decaycon}
f_{a_0^-}(\mu)=\bar f_{a_0}\,{m_d(\mu)-m_u(\mu)\over m_{a_0} }, \qquad
f_{a_0^+}(\mu)=\bar f_{a_0}\,{m_u(\mu)-m_d(\mu)\over m_{a_0} }.
\en
Therefore, the vector decay constants of  $a_0^-$ and $a_0^+$ are of opposite sign.

In \cite{CCY:SP} we have applied the QCD sum rule method to
estimate the decay constant $\bar f_S$ for various scalar mesons as summarized in Table \ref{tab:moments}. Note that a recent sum rule calculation in \cite{Han} yields a smaller $\bar f_S$ in scenario 2 for $S=f_0(1710),a_0(1450)$ and $K_0^*(1430)$.
In this work we shall use the values of $f_V$ and $f_V^\bot$ taken from \cite{BallfV}.
For the decay constants $f^{q}_{\eta^{(')}}$ and $f_{\eta^{(')}}^{s}$ of the $\eta$ and $\eta'$ mesons defined by
\be
\la 0|\bar q\gamma_\mu\gamma_5q|\eta^{(')}\ra=i{1\over\sqrt{2}}f_{\eta^{(')}}^q  p_\mu, \quad \la 0|\bar s\gamma_\mu\gamma_5s|\eta^{(')}\ra=if_{\eta^{(')}}^s p_\mu,
\en
we shall follow the results of \cite{BenekeETA}.

For the  $B\to P$ and $B\to V$ transition form factors defined in the conventional way \cite{BSW}, we will use the results obtained using the QCD sum rule method \cite{Ball}. Form factors for $B\to S$ transitions are defined by \cite{CCH}
 \be \label{eq:FF}
\la S(p')|A_\mu|B(p)\ra &=& -i\Bigg[\left(P_\mu-{m_B^2-m_S^2\over
q^2}\,q_ \mu\right) F_1^{BS}(q^2)   +{m_B^2-m_S^2\over
q^2}q_\mu\,F_0^{BS}(q^2)\Bigg],
 \en
where $P_\mu=(p+p')_\mu$ and $q_\mu=(p-p')_\mu$.
The momentum dependence of the form factor is usually parameterized in a
3-parameter form
\be
F(q^2)={F(0)\over 1-a(q^2/m_B^2)+b(q^4/m_B^4)}.
\en
The parameters $F(0)$, $a$ and $b$ for $B\to S$ transitions are summarized in Table \ref{tab:FF}
obtained using the covariant light-front quark model \cite{CCH}. Form factors are also available in other approaches, such as light-cone sum rule \cite{Aliev,Lu:2008,Wang,Sun,Han} and pQCD \cite{Lu:2009,Xiao:K0steta}. In general, form factors calculated by sum rule and pQCD methods are larger than that obtained using the quark model. For example, $F_0^{BK_0^*}(0)$ is of order 0.26 in the covariant light-front quark model \cite{CCH}, while it is found to be 0.45 \cite{Han}, 0.49 \cite{Lu:2008,Sun} in the sum rule method and 0.60 \cite{Lu:2009} and 0.76 \cite{Xiao:K0steta} in pQCD (all evaluated in scenario 2). We will come to this point later.

\subsection{Distribution amplitudes}
In general, the twist-2 light-cone distribution amplitude (LCDA) of the scalar meson $\Phi_S$
has the form
 \be \label{eq:twist2wf}
 \Phi_S(x,\mu)=f_S\,6x(1-x)\left[1+\mu_S\sum_{m=1}^\infty
 B_m(\mu)\,C_m^{3/2}(2x-1)\right],
 \en
where $B_m$ are Gegenbauer moments and $C_m^{3/2}$ are
Gegenbauer polynomials. The general  twist-3 LCDAs are given by
\be \label{eq:twist3wf}
 \Phi^s_S(x) &=&\bar f_S\left[ 1+\sum_{m=1}^\infty a_m(\mu)C_m^{1/2}(2u-1)\right], \non \\
  \Phi^\sigma_S(x) &=& \bar f_S \,
 6x(1-x)\left[ 1+\sum_{m=1}^\infty b_m(\mu)C_m^{1/2}(2u-1)\right].
 \en
Since $\mu_S\equiv 1/B_0\gg 1$ and even
Gegenbauer coefficients $B_m$ are suppressed, it is clear that the twist-2 LCDA of the scalar meson is dominated by the odd Gegenabuer moments. In
contrast, the odd Gegenbauer moments vanish for the $\pi$ and
$\rho$ mesons. The Gegenbauer moments $B_1$ and $B_3$   in scenarios 1
and 2  obtained using the QCD sum rule method \cite{CCY:SP} are listed in Table \ref{tab:moments}. The Gegenbauer moments $a_{1,2,4}$ and $b_{1,2,4}$  for twist-3 LCDAs have been computed in \cite{Lu:2007,Han}.

Since the decay constants vanish for the neutral scalar mesons $f_0,~a_0^0$ and $\sigma$, it follows from Eq. (\ref{eq:twist2wf}) that
\be \label{}
 \Phi_S(x,\mu)=\bar f_S 6x(1-x)\sum_{m=1}^\infty
 B_m(\mu)\,C_m^{3/2}(2x-1)
\en
for these neutral scalar mesons.

As stressed in \cite{CCY:SP}, it is most suitable to define the LCDAs
of scalar mesons including decay constants. However,  it is
more convenient in practical calculations to factor out the decay constants in the LCDAs and put them back in the appropriate places. In the ensuing
discussions, we will use the LCDAs with the decay constants
$f_S,\bar f_S,f_V, f_V^\bot,f_P$ being factored out.

\subsection{Mixing angle between $f_0(980)$ and $f_0(500)$ and between $\eta$ and $\eta'$}

In the naive 2-quark model with ideal mixing for $f_0(980)$ and $f_0(500)$, $f_0(980)$ is purely an $s\bar s$ state, while $f_0(500)$ is a $n\bar n$ state with $n\bar n\equiv (\bar uu+\bar dd)/\sqrt{2}$.
However, there also exist
some experimental evidences indicating that $f_0(980)$ is not
purely an $s\bar s$ state. For example, the observation of
$\Gamma(J/\psi\to f_0\omega)\approx {1\over 2}\Gamma(J/\psi\to
f_0\phi)$ \cite{PDG} clearly shows the existence of the
non-strange and strange quark content in $f_0(980)$.  Therefore, isoscalars $f_0(500)$ and $f_0(980)$ must have a mixing
\be \label{eq:f0mixing}
   \left( \begin{array}{c}
    |f_0(980)\rangle \\ |f_0(500)\rangle
   \end{array} \right)
   = \left( \begin{array}{ccc}
    \cos\theta & \sin\theta \\
    -\sin\theta & \cos\theta
   \end{array} \right)
   \left( \begin{array}{c}
    |n\bar n\rangle \\ |s\bar s\rangle
   \end{array} \right) \;.
\en
Various mixing angle measurements have been discussed in the literature and summarized in \cite{CCY:SP,Fleischer:2011au}. A recent measurement of the upper limit on the branching fraction product $\B(\ov B^0\to J/\psi f_0(980))\times\B(f_0(980)\to \pi^+\pi^-)$ by LHCb leads to $|\theta|<30^\circ$ \cite{LHCb:theta}.

For the $\eta$ and $\eta'$ mesons,
it is more convenient to consider the flavor states   $q\bar q\equiv (u\bar u+d\bar
d)/\sqrt{2}$, $s\bar s$ and $c\bar c$ labeled by the $\eta_q$, $\eta_s$ and $\eta_{c}^0$, respectively. Neglecting the small mixing with $\eta_c^0$, we write
\begin{equation}\label{eq:qsmixing}
   \left( \begin{array}{c}
    |\eta\rangle \\ |\eta'\rangle
   \end{array} \right)
   = \left( \begin{array}{ccc}
    \cos\phi & -\sin\phi \\
    \sin\phi & \cos\phi
   \end{array} \right)
   \left( \begin{array}{c}
    |\eta_q\rangle \\ |\eta_s\rangle
   \end{array} \right) \;,
\end{equation}
where $\phi=(39.3\pm1.0)^\circ$ \cite{FKS} is the $\eta-\eta'$ mixing angle in the $\eta_q$ and $\eta_s$ flavor basis.

\section{Decay amplitudes in QCD factorization}
 We shall use the QCD factorization approach
\cite{BBNS,BN} to study the short-distance contributions to the
$B\to SP,SV$ decays with $S=f_0(980),a_0(980),a_0(1450),K^*_0(1430)$. In QCD factorization, the factorizable amplitudes of above-mentioned decays
can be found in \cite{CCY:SP} and \cite{CCY:SV}. However, the expressions for the decay amplitudes of $B\to
f_0K,~a_0^0\pi,~a_0^0K$ involving a neutral $f_0$ or $a_0$ given in \cite{CCY:SP} are corrected in Appendix A as some factorizable contributions were missed before.
The effective parameters $a_i^p$ with $p=u,c$ appearing in
Eq. (\ref{eq:SDAmp}) can be calculated in the QCD factorization
approach \cite{BBNS}. In general, they have the expressions
\be  \label{eq:ai}
  a_i^p(M_1M_2) =
 \left(c_i+{c_{i\pm1}\over N_c}\right)N_i(M_2)
   + {c_{i\pm1}\over N_c}\,{C_F\alpha_s\over
 4\pi}\Big[V_i(M_2)+{4\pi^2\over N_c}H_i(M_1M_2)\Big]+P_i^p(M_2),
\en
where $i=1,\cdots,10$,  the upper (lower) signs apply when $i$ is
odd (even), $c_i$ are the Wilson coefficients,
$C_F=(N_c^2-1)/(2N_c)$ with $N_c=3$, $M_2$ is the emitted meson
and $M_1$ shares the same spectator quark with the $B$ meson. The
quantities $V_i(M_2)$ account for vertex corrections,
$H_i(M_1M_2)$ for hard spectator interactions with a hard gluon
exchange between the emitted meson and the spectator quark of the
$B$ meson and $P_i(M_2)$ for penguin contractions. The expression
of the quantities $N_i(M_2)$ reads
 \be
 N_i(M_2)=\cases{0, & $i=6,8$~and~$M_2=V$, \cr
                 1, & else. \cr}
 \en
The explicit expressions of $V_i(M)$, $H_i(M_1M_2)$, and weak annihilation contributions described by the terms $b_i$ and $b_{i,\rm EW}$ are given in \cite{CCY:SP} and \cite{CCY:SV} for $B\to SP$ and $B\to SV$, respectively. \footnote{In Eq. (4.8) of \cite{CCY:SP}, the second term of the annihilation amplitude $A_3^f$ for $M_1M_2=SP$ should have an identical expression, including the sign, as that for $M_1M_2=PS$. Note that the expression was correct in the original archive version, arXiv:hep-ph/0508104.}

Power corrections in QCDF always involve troublesome endpoint divergences.  We shall follow \cite{BBNS} to model the endpoint divergence $X\equiv\int^1_0 dx/(1-x)$ in the annihilation and hard spectator
scattering diagrams as
 \be \label{eq:XA}
 X_A=\ln\left({m_B\over \Lambda_h}\right)(1+\rho_A e^{i\phi_A}), \qquad
 X_H=\ln\left({m_B\over \Lambda_h}\right)(1+\rho_H e^{i\phi_H}),
 \en
with $\Lambda_h$ being a typical scale of order
500 MeV, and  $\rho_{A,H}$, $\phi_{A,H}$ being the unknown real parameters.

In principle, physics should
be independent of the choice of $\mu$, but in practice there exists some
residual $\mu$ dependence in the truncated calculations. However, we found that sometimes even the decay rates without annihilation are sensitive to the choice of $\mu$. For example, we found that the measured branching fractions of $B\to K_0^*(1430)(\eta,\eta')$ cannot be accommodated for $\mu=m_b/2$. Indeed, this observation also occurs in our previous study of $B\to VV$ decays \cite{ChengVV}.  We found that if the renormalization scale is chosen to be $\mu=m_b(m_b)/2=2.1$ GeV, we cannot fit the branching fractions and polarization
fractions simultaneously for both $B\to K^*\phi$ and $B\to K^*\rho$ decays. Therefore, we will confine
ourselves to the renormalization scale $\mu=m_b(m_b)$ in the ensuing study.
Note that the hard spectator and annihilation contributions should be evaluated at the hard-collinear scale $\mu_h=\sqrt{\mu\Lambda_h}$ with $\Lambda_h\approx 500$ MeV \cite{BBNS}.

As discussed in \cite{CCY:SP} and \cite{CCY:SV}, scenario 2 in which the scalar mesons above 1 GeV are lowest lying $q\bar q$ scalar state and the light scalar mesons are four-quark states is preferable, while all scalar mesons are made of $q\bar q$ quarks in scenario 1. It is widely believed that the $f_0(980)$ and the $a_0(980)$ are predominately four-quark states, but in practice it is difficult to make quantitative predictions on hadronic $B\to SP,SV$ decays
based on the four-quark picture for light scalar mesons.
Hence, we shall assume scenario 1 for the $f_0(980)$
and the $a_0(980)$ in order to apply QCDF.

\begin{table}[!]
\caption{Branching fractions (in units of $10^{-6}$) of $B$ decays to
a scalar meson and a pseudoscalar meson.  The predicted rates of
$B\to f_0(980)K,f_0(980)\pi$ are for the $f_0(980)-f_0(500)$ mixing angle
$\theta=17^\circ$. We work in scenario 1 for the light scalar mesons $f_0(980)$ and $a_0(980)$ and scenario 2 for the scalar mesons $a_0(1450)$ and $K_0^*(1430)$; see the main text for explanation. Experimental results are taken from Table \ref{tab:exptBR}. We have used $\B(f_0(980)\to \pi^+\pi^-)=0.50$ and $\B(a_0(980)\to \pi\eta)=0.845\pm0.017$ to obtain the experimental branching fractions for $f_0(980)P$ and $a_0(980)P$. For comparison, predictions based on the pQCD approach are also exhibited.}
\label{tab:BRSP}
\begin{ruledtabular}
\footnotesize{
\begin{tabular}{l c c c }
Mode & QCDF (this work) & pQCD \cite{Xiao:K0steta,Zhang:a0pi,Zhang:f0pi,WangWei,Shen} & Expt   \\
\hline
 $B^-\to f_0(980)K^-$ & $16.1^{+1.9+1.2+30.8}_{-1.8-1.8-11.0}$ & $16\sim 18$ \footnotemark[1] & $18.8^{+1.8}_{-2.0}$ \\
 $\ov B^0\to f_0(980)\ov K^0$ & $14.8^{+1.7+1.1+28.6}_{-1.6-1.6-10.2}$  & $13\sim 16$ \footnotemark[1] & $14.0\pm1.8$  \\
 $B^-\to f_0(980)\pi^-$ & $0.26^{+0.04+0.05+0.18}_{-0.03-0.04-0.12}$ & $2.5\pm1.0$ & $<3.0$ \\
 $\ov B^0\to f_0(980)\pi^0$ & $0.08^{+0.01+0.01+0.08}_{-0.01-0.01-0.03}$  & $0.26\pm0.06$  & \\
  $B^-\to a_0^0(980)K^-$                                      &
                                                                 $0.34^{+0.08+0.45+1.02}_{-0.06-0.13-0.07}$ &$3.5^{+0.4+0.4+1.0}_{-0.4-0.6-1.0}$ & $<3.0$  \\
  $B^-\to a_0^-(980)\ov K^0$
                                                                 &   $0.08^{+0.12+0.58+2.12}_{-0.07-0.08-0.02}$ &$6.9^{+0.8+1.1+2.0}_{-0.7-1.1-1.7}$ & $<4.6$ \\
  $\bar B^0\to a_0^+(980)K^-$
                                                                 & $0.34^{+0.05+0.62+2.27}_{-0.05-0.13-0.00}$
                                                                 &$9.7^{+1.1+1.6+2.7}_{-1.0-1.4-2.2}$ & $<2.2$  \\
  $\bar B^0\to a_0^0(980)\ov K^0$                                    &
                                                                 $0.05^{+0.05+0.28+0.86}_{-0.03-0.04-0.00}$ &$4.7^{+0.5+0.7+1.1}_{-0.5-0.8-1.1}$ & $<9.2$ \\

 $B^-\to a_0^0(980)\pi^-$ & $4.9^{+0.4+1.2+0.5}_{-0.3-1.1-0.7}$ & $2.8^{+0.0+0.0+0.0}_{-0.8-0.9-0.6}$ & $<6.9$\\
 $B^-\to a_0^-(980)\pi^0$ & $0.70^{+0.22+0.09+0.22}_{-0.16-0.08-0.14}$ & $0.41^{+0.00+0.00+0.00}_{-0.13-0.14-0.12}$ & $<1.7$\\
 $\ov B^0\to a_0^+(980)\pi^-$
 & $5.3^{+0.3+1.4+1.0}_{-0.4-1.3-0.4}$ & $0.86^{+0.10+0.14+0.01}_{-0.09-0.14-0.00}$ &  \\
 $\ov B^0\to a_0^-(980)\pi^+$
 & $0.58^{+0.11+0.09+0.63}_{-0.09-0.08-0.22}$ & $0.51^{+0.05+0.09+0.07}_{-0.06-0.09-0.06}$ &   \\
 $\ov B^0\to a_0^\pm(980)\pi^\mp$
 & $4.7^{+0.6+1.3+1.8}_{-0.6-1.1-0.5}$ & $0.93^{+0.10+0.15+0.02}_{-0.10-0.14-0.00}$ & $<3.7$ \\
 $\ov B^0\to a_0^0(980)\pi^0$ &
 $1.0^{+0.5+0.1+0.2}_{-0.3-0.1-0.1}$ & $0.51^{+0.08+0.09+0.00}_{-0.07-0.09-0.00}$ &  \\
 $B^-\to a_0^0(1450)K^-$ & $2.2^{+0.7+2.3+7.7}_{-0.5-1.0-1.9}$ &  & \\
 $B^-\to a_0^-(1450)\ov K^0$ & $4.2^{+1.6+4.9+18.1}_{-1.2-2.1-4.2}$ &  & \\
 $\ov B^0\to a_0^+(1450)K^-$ &
 $3.5^{+1.0+4.5+16.9}_{-0.8-1.9-~3.3}$ &  & $<4.7$ \\
 $\ov B^0\to a_0^0(1450)\ov K^0$
 & $1.9^{+0.8+2.1+7.5}_{-0.6-0.9-1.9}$ &   \\
 $B^-\to a_0^0(1450)\pi^-$ & $5.1^{+0.5+1.2+1.3}_{-0.4-1.1-1.3}$ & & \\
 $B^-\to a_0^-(1450)\pi^0$ & $2.1^{+0.7+0.2+0.8}_{-0.5-0.2-0.6}$ &  & \\
 $\ov B^0\to a_0^+(1450)\pi^-$
 & $2.5^{+0.5+0.9+3.7}_{-0.5-0.8-0.3}$ &  &  \\
 $\ov B^0\to a_0^-(1450)\pi^+$
 & $0.74^{+0.20+0.19+2.92}_{-0.16-0.17-0.51}$ &   \\
 $\ov B^0\to a_0^\pm(1450)\pi^\mp$
 & $1.3^{+0.7+0.7+8.3}_{-0.6-0.6-0.1}$ &  & $<3.5$  \\
 $\ov B^0\to a_0^0(1450)\pi^0$ &
 $3.3^{+1.4+0.3+2.8}_{-1.0-0.3-1.3}$ &  \\
   $B^-\to \bar K_0^{*0}(1430)\pi^-$                                               &
                                                                 $12.9^{+4.6+4.1+38.5}_{-3.7-3.4-9.1}$ &$30.9^{+12.5}_{-~9.2}$ \footnotemark[2] & $45.1\pm6.3$  \\
  $B^-\to K_0^{*-}(1430)\pi^0$
                                                                 &   $7.4^{+2.4+2.1+20.1}_{-1.9-1.8-~5.0}$ &$21.6^{+8.5}_{-6.6}$ \footnotemark[2] & $$ \\
  $\bar B^0\to K_0^{*-}(1430)\pi^+$
                                                                 & $13.8^{+4.5+4.1+38.3}_{-3.6-3.5-~9.5}$
                                                                 &$31.6^{+12.4}_{-~9.3}$ \footnotemark[2] & $33.5^{+3.9}_{-3.8}$  \\
  $\bar B^0\to \bar K_0^{*0}(1430)\pi^0$                                    &
                                                                 $5.6^{+2.6+2.4+18.8}_{-1.3-1.2-~3.9}$ &$10.7^{+4.1}_{-3.2}$ \footnotemark[2] & $7.0\pm1.2$ \\
  $B^-\to K_0^{*-}(1430)\eta$                                               &
                                                                 $17.9^{+3.9+8.3+~9.1}_{-3.4-5.3-12.3}$ &$33.8^{+13.5+1.1+7.7+8.2}_{-~9.0-1.1-7.0-7.3}$ & $15.8\pm3.1$  \\
  $B^-\to K_0^{*-}(1430)\eta'$
                                                                 &   $9.3^{+4.7+4.0+51.6}_{-3.6-4.4-~8.0}$ &$77.5^{+15.8+6.2+21.0+18.0}_{-10.8-5.8-16.5-16.1}$ & $5.2\pm2.1$ \\
  $\bar B^0\to \bar K_0^{*0}(1430)\eta$
                                                                 & $16.1^{+3.6+7.6+~9.1}_{-3.1-4.9-11.7}$
                                                                 &$28.4^{+11.6+1.4+6.4+6.9}_{-~7.8-1.4-5.9-6.2}$ & $9.6\pm1.9$  \\
  $\bar B^0\to \bar K_0^{*0}(1430)\eta'$                                    &
                                                                 $8.7^{+4.4+3.7+48.7}_{-3.3-4.1-~7.5}$ &$74.2^{+15.0+6.4+20.5+17.2}_{-10.3-5.7-16.2-15.5}$ & $6.3\pm1.6$ \\
\end{tabular}
}
\end{ruledtabular}
\footnotetext[1]{For the mixing angle $140^\circ<\theta<165^\circ$ \cite{WangWei}.}
\footnotetext[2]{Results based on the new Gegenbauer moments obtained in \cite{Xiao:K0steta}. For previous pQCD calculations of $B\to K^*_0(1430)\pi$, see \cite{Shen}.}
\end{table}

\begin{table}[!]
\caption{Same as Table \ref{tab:BRSP} except for $B$ decays to
a scalar meson and a vector meson.}
\label{tab:BRSV}
\begin{ruledtabular}
\begin{tabular}{l c c c}
Mode & QCDF (this work) & pQCD \cite{Zhang:a01450rho,Zhang:a01450Kst,Zhang:f0Kst,Zhang:K0strho,CSKim}  & Expt  \\
\hline
 $B^-\to f_0(980)K^{*-}$ & $10.7^{+0.3+0.2+4.8}_{-0.3-0.2-0.5}$ & $11.7\sim 14.6$ & $11.4\pm1.4$\\
 $\ov B^0\to f_0(980)\ov K^{*0}$ & $9.1^{+1.0+1.0+5.3}_{-0.4-0.5-0.7}$  & $11.2\sim 13.7$ & $8.4\pm1.4$  \\
 $B^-\to f_0(980)\rho^-$ & $0.35^{+0.01+0.08+0.09}_{-0.01-0.07-0.08}$ & $0.7^{+0.1+0.2+0.2}_{-0.0-0.1-0.1}$ & $<4.0$ \\
 $\ov B^0\to f_0(980)\rho^0$ & $0.02^{+0.00+0.00+0.03}_{-0.00-0.00-0.00}$  & $0.33^{+0.04+0.07+0.06}_{-0.03-0.05-0.06}$ & $1.7\pm0.6$  \\
 $\ov B^0\to f_0(980)\omega$ & $0.02^{+0.00+0.00+0.03}_{-0.00-0.00-0.01}$  & $0.34^{+0.03+0.06+0.06}_{-0.04-0.06-0.05}$ & $<3.0$   \\
 $B^-\to a_0^0(980)K^{*-}$ & $2.4^{+0.3+0.4+2.8}_{-0.3-0.4-0.0}$ &  &\\
 $B^-\to a_0^-(980)\ov K^{*0}$ & $8.4^{+1.3+1.2+8.2}_{-1.1-1.1-0.2}$ &  & \\
 $\ov B^0\to a_0^+(980)K^{*-}$ &
 $5.6^{+1.2+1.4+7.9}_{-0.2-0.4-0.6}$ & &  \\
 $\ov B^0\to a_0^0(980)\ov K^{*0}$
 & $3.2^{+2.5+2.7+2.8}_{-2.9-2.7-0.0}$ &  &  \\
 $B^-\to a_0^0(980)\rho^-$ & $6.8^{+0.6+1.7+0.9}_{-0.5-1.5-0.3}$ &  & \\
 $B^-\to a_0^-(980)\rho^0$ & $2.1^{+0.6+0.0+0.9}_{-0.5-0.0-0.2}$ &  & \\
 $\ov B^0\to a_0^+(980)\rho^-$
 & $22.5^{+4.0+6.8+5.0}_{-0.1-2.7-2.3}$ &   \\
 $\ov B^0\to a_0^-(980)\rho^+$
 & $0.60^{+0.14+0.05+0.47}_{-0.03-0.05-0.07}$ &   \\
 $\ov B^0\to a_0^0(980)\rho^0$ &
 $1.3^{+0.8+0.1+0.4}_{-0.4-0.1-0.0}$ &  \\
 $B^-\to a_0^-(980)\omega$ & $1.0^{+0.4+0.0+0.5}_{-0.3-0.0-0.2}$ & & \\
 $\ov B^0\to a_0^0(980)\omega$ &
 $0.4^{+0.3+0.0+0.2}_{-0.1-0.0-0.0}$ &  \\
 $B^-\to a_0^0(1450)K^{*-}$ & $2.5^{+0.3+0.4+5.6}_{-0.3-0.4-0.1}$ &  $7.0^{+0.9+1.6+1.7+0.2}_{-0.7-1.1-1.4-0.0}$ & \\
 $B^-\to a_0^-(1450)\ov K^{*0}$ & $8.4^{+1.4+1.3+14.6}_{-1.1-1.2-~0.4}$ & $3.0^{+0.2+0.2+0.7+1.2}_{-0.1-0.1-0.6-0.7}$ & \\
 $\ov B^0\to a_0^+(1450)K^{*-}$ &
 $8.9^{+1.9+3.0+3.9}_{-0.3-1.4-0.5}$ & $2.8^{+0.3+0.1+0.7+0.8}_{-0.3-0.0-0.5-0.6}$  \\
 $\ov B^0\to a_0^0(1450)\ov K^{*0}$
 & $3.4^{+0.8+0.9+6.7}_{-0.2-0.3-0.3}$ & $1.4^{+0.1+0.0+0.3+0.5}_{-0.1-0.1-0.3-0.4}$  \\
 $B^-\to a_0^0(1450)\rho^-$ & $3.7^{+0.6+1.4+1.9}_{-0.5-1.1-0.3}$ & $79.3^{+33.2+18.2+5.5+3.9}_{-22.1-16.3-4.7-3.5}$ & \\
 $B^-\to a_0^-(1450)\rho^0$ & $3.2^{+0.8+0.2+2.3}_{-0.6-0.2-0.4}$ & $1.9^{+0.9+0.4+0.1+0.4}_{-0.5-0.4-0.0-0.3}$ & \\
 $\ov B^0\to a_0^+(1450)\rho^-$
 & $11.2^{+2.4+4.5+11.2}_{-0.4-2.3-~1.0}$ &  $184.3^{+69.3+42.2+13.0+13.7}_{-47.9-37.9-13.3-14.1}$ \\
 $\ov B^0\to a_0^-(1450)\rho^+$
 & $1.2^{+0.3+0.1+1.3}_{-0.1-0.1-0.1}$ & $3.6^{+0.7+0.8+0.8+0.4}_{-0.6-0.8-0.8-0.3}$   \\
 $\ov B^0\to a_0^0(1450)\rho^0$ &
 $4.4^{+2.7+0.5+1.1}_{-1.3-0.2-1.2}$ & $7.2^{+2.1+1.7+1.1+1.6}_{-1.6-1.3-0.9-1.3}$  \\
 $B^-\to a_0^-(1450)\omega$ & $1.5^{+0.6+0.0+1.5}_{-0.4-0.0-0.3}$ & $0.3^{+0.4+0.1+0.1+0.1}_{-0.1-0.0-0.0-0.0}$ & \\
 $\ov B^0\to a_0^0(1450)\omega$ &
 $1.8^{+1.1+0.2+1.7}_{-0.5-0.1-0.3}$ & $2.1^{+0.7+0.4+0.2+0.1}_{-0.6-0.5-0.2-0.2}$  \\
  $B^-\to \bar K_0^{*0}(1430)\rho^-$                                               &
                                                                 $39.0^{+34.5+35.2+41.6}_{-35.8-35.3-51.0}$ &$12.1^{+2.8+3.9+0.5}_{-0.0-3.1-0.5}$ & $$  \\
  $B^-\to K_0^{*-}(1430)\rho^0$
                                                                 &   $14.8^{+3.7+0.4+6.7}_{-3.2-0.4-0.2}$ &$8.4^{+2.3+3.3+0.9}_{-0.0-3.2-0.7}$ & $$ \\
  $\bar B^0\to K_0^{*-}(1430)\rho^+$
                                                                 & $36.3^{+8.5+0.8+16.7}_{-7.4-0.8-~0.1}$
                                                                 &$10.5^{+2.7+3.5+0.3}_{-0.0-2.6-0.3}$ &$28.0\pm11.6$  \\
  $\bar B^0\to \bar K_0^{*0}(1430)\rho^0$                                    &
                                                                 $23.4^{+5.1+0.6+9.4}_{-4.5-0.5-0.4}$ &$4.8^{+1.1+1.0+0.3}_{-0.0-1.0-0.3}$ & $27.0\pm5.4$ \\
 $B^-\to K_0^{*-}(1430)\omega$                                    &
                                                                 $21.5^{+5.8+0.5+9.8}_{-4.9-0.5-3.4}$ &$7.4^{+2.1+3.0+0.9}_{-1.5-2.3-0.4}$ & $24.0\pm5.1$ \\
 $\bar B^0\to \bar K_0^{*0}(1430)\omega$                                    &
                                                                 $21.9^{+5.9+0.6+10.5}_{-5.0-0.6-~3.3}$ &$9.3^{+2.1+3.6+1.2}_{-2.0-2.9-1.0}$ & $16.0\pm3.4$ \\
 $B^-\to K_0^{*-}(1430)\phi$                                    &
                                                                 $3.8^{+0.7+0.1+11.2}_{-0.6-0.1-~1.6}$ &$25.6^{+6.2+0.9+12.1}_{-5.4-0.8-~6.5}$ & $7.0\pm1.6$ \\
 $\bar B^0\to \bar K_0^{*0}(1430)\phi$                                    &
                                                                 $3.7^{+0.8+0.1+5.6}_{-0.6-0.1-0.1}$ &$23.6^{+5.6+0.8+10.9}_{-5.0-0.6-~5.8}$ & $3.9\pm0.8$ \\
\end{tabular}
\end{ruledtabular}
\end{table}


\begin{table}[t]
\caption{Branching fractions (in units of $10^{-6}$) of $B\to SP$ (upper) and $B\to SV$ (lower) decays with $S=a_0(1450),K_0^*(1430)$ in QCD factorization.   Experimental results are taken from Table \ref{tab:exptBR}. The scalar mesons
$a_0(1450)$ and $K_0^*(1430)$ are treated as the first excited states of low lying light $q\bar q$ scalars $a_0(980)$ and $K_0^*(800)$, corresponding to scenario 1 as explained in the main text.} \label{tab:theoryBRS1}
\begin{ruledtabular}
\begin{tabular}{l r c |l  r c}
Mode & Theory & Expt & Mode & Theory & Expt  \\
\hline
 $B^-\to a_0^0(1450)K^-$
 & $0.7^{+0.7+0.3+3.1}_{-0.4-0.1-0.5}$ &  & $\ov B^0\to a_0^+(1450)K^-$
 & $1.9^{+1.9+0.9+7.8}_{-1.1-0.4-1.4}$ & $<4.7$ \non \\
 $B^-\to a_0^-(1450)\ov K^0$
 & $2.7^{+2.8+1.2+9.6}_{-1.7-0.6-2.7}$ &  & $\ov B^0\to a_0^0(1450)\ov K^0$
 & $0.9^{+1.0+0.4+3.6}_{-0.6-0.2-0.9}$ \non \\
 $B^-\to a_0^0(1450)\pi^-$
 & $2.7^{+0.1+0.6+0.4}_{-0.1-0.5-0.5}$ &  & $\ov B^0\to a_0^+(1450)\pi^-$
 & $11.2^{+2.0+1.8+4.5}_{-1.6-1.7-5.2}$ &  \non \\
 $B^-\to a_0^-(1450)\pi^0$
 & $0.4^{+0.2+0.0+0.2}_{-0.2-0.0-0.2}$ & & $\ov B^0\to a_0^-(1450)\pi^+$
 & $0.02^{+0.05+0.01+0.75}_{-0.00-0.00-0.01}$ \non \\
 & & &  $\ov B^0\to a_0^\pm(1450)\pi^\mp$ &
 $11.9^{+2.7+1.8+5.6}_{-2.3-1.6-3.7}$ & $<3.5$ \non \\
 & & &  $\ov B^0\to a_0^0(1450)\pi^0$ &
 $1.3^{+0.7+0.1+2.0}_{-0.5-0.1-1.0}$ & \non \\
 $B^-\to \ov K^{*0}_0(1430)\pi^-$
 & $1.3^{+0.7+1.0+15.9}_{-0.5-0.7-~1.1}$ &  $45.1\pm6.3$ & $\ov B^0\to K^{*-}_0(1430)\pi^+$
 & $1.6^{+0.8+1.0+15.7}_{-0.6-0.8-~1.3}$ & $32.1\pm3.7$ \non \\
 $B^-\to K^{*-}_0(1430)\pi^0$
 & $0.3^{+0.2+0.3+6.4}_{-0.1-0.2-0.3}$ & &  $\ov B^0\to \ov K^{*0}_0(1430)\pi^0$
 & $1.3^{+0.7+0.8+9.4}_{-0.3-0.5-1.3}$ & $7.0\pm1.2$ \non \\
 $B^-\to K_0^{*-}(1430)\eta$ & $5.4^{+1.9+1.2+3.7}_{-1.6-1.0-4.4}$ & $15.8\pm3.1$  & $\ov B^0\to\ov K_0^{*0}\eta$ & $5.8^{+2.1+1.4+4.9}_{-1.7-1.1-5.2}$ & $9.6\pm1.9$ \non \\
 $B^-\to K_0^{*-}(1430)\eta'$ & $6.2^{+1.0+1.0+35.8}_{-0.8-1.0-~6.1}$ & $5.2\pm2.1$ & $\ov B^0\to\ov K_0^{*0}\eta'$ & $5.9^{+0.9+0.9+33.6}_{-0.8-0.9-5.8}$ & $6.3\pm1.6$ \non \\
 \hline
 $B^-\to a_0^0(1450)K^{*-}$ & $0.7^{+0.1+0.2+1.7}_{-0.1-0.2-0.0}$ &  & $\ov B^0\to a_0^+(1450)K^{*-}$ &
 $4.8^{+0.9+2.1+7.3}_{-0.1-1.1-0.4}$ &  \non \\
 $B^-\to a_0^-(1450)\ov K^{*0}$ & $1.3^{+0.2+0.5+4.2}_{-0.1-0.4-0.2}$ &  & $\ov B^0\to a_0^0(1450)\ov K^{*0}$
 & $0.6^{+0.1+0.3+1.9}_{-0.0-0.2-0.1}$ &  \non \\
 $B^-\to a_0^0(1450)\rho^-$ & $6.9^{+1.3+1.8+0.2}_{-1.1-1.6-1.1}$ & & $\ov B^0\to a_0^+(1450)\rho^-$
 & $17.3^{+4.2+5.6+2.3}_{-1.0-2.4-6.0}$ &  \non \\
 $B^-\to a_0^-(1450)\rho^0$ & $0.2^{+0.1+0.0+0.6}_{-0.1-0.0-0.1}$ &  & $\ov B^0\to a_0^-(1450)\rho^+$
 & $0.2^{+0.1+0.0+0.3}_{-0.1-0.0-0.0}$ &  \non \\
 $B^-\to a_0^-(1450)\omega$ & $0.5^{+0.1+0.1+0.4}_{-0.1-0.1-0.1}$ & &  $\ov B^0\to a_0^0(1450)\rho^0$ &
 $1.4^{+1.6+0.2+0.3}_{-0.8-0.1-0.8}$ & \non \\
 & & &  $\ov B^0\to a_0^0(1450)\omega$ &
 $0.1^{+0.1+0.0+0.5}_{-0.0-0.0-0.0}$ & \non \\
 $B^-\to K^{*-}_0(1430)\phi$ &  $2.8^{+0.5+0.0+5.0}_{-0.4-0.0-0.5}$ & $7.0\pm1.6$
 & $\ov B^0\to \ov K^{*0}_0(1430)\phi$ &
 $2.5^{+0.5+0.0+4.7}_{-0.3-0.0-0.5}$& $3.9\pm0.8$ \non \\
 $B^-\to \ov K^{*0}_0(1430)\rho^-$ & $13.2^{+~9.9+10.4+10.8}_{-10.8-10.4-15.1}$ & $$ &
 $\ov B^0\to K^{*-}_0(1430)\rho^+$  & $10.9^{+2.5+0.3+5.5}_{-2.2-0.3-0.0}$ & $28.0\pm11.6$ \non \\
 $B^-\to K^{*-}_0(1430)\rho^0$ & $8.4^{+1.7+0.2+2.1}_{-1.5-0.2-0.1}$ &  &
 $\ov B^0\to \ov K^{*0}_0(1430)\rho^0$  & $4.1^{+1.1+0.2+2.6}_{-1.0-0.2-0.1}$ &
 $27.0\pm5.4$  \non \\
 $B^-\to K^{*-}_0(1430)\omega$ & $10.6^{+3.0+0.2+3.8}_{-2.4-0.2-1.4}$ & $24.0\pm5.1$ &
 $\ov B^0\to \ov K^{*0}_0(1430)\omega$  & $9.3^{+2.7+0.3+3.9}_{-2.2-0.3-1.3}$ &
 $16.0\pm3.4$  \non \\
\end{tabular}
\end{ruledtabular}
\end{table}

\begin{table}[t]
\caption{{\it CP} asymmetries (in units of \%) of $B$ decays to
a scalar meson and a pseudoscalar meson.   We work in scenario 1 for the light scalar mesons $f_0(980)$ and $a_0(980)$ and scenario 2 for the scalar mesons $a_0(1450)$ and $K_0^*(1450)$. Experimental results are taken from Table \ref{tab:exptCP}. }
\label{tab:asySP}
\begin{ruledtabular}
\footnotesize{
\begin{tabular}{l r c |l  r c}
Mode & Theory & Expt & Mode & Theory & Expt  \\
\hline
 $B^-\to f_0(980)K^-$ & $1.4^{+0.2+0.5+1.1}_{-0.2-0.4-1.4}$ & $-9.5^{+4.9}_{-4.2}$ &
 $\ov B^0\to f_0(980)\ov K^0$ & $2.9^{+0.4+0.5+4.8}_{-0.3-0.2-2.6}$  & $-14\pm12$ \footnotemark[1]\non \\
 $B^-\to f_0(980)\pi^-$ & $19.0^{+3.1+2.0+36.1}_{-2.6-2.4-28.4}$ &  &
 $\ov B^0\to f_0(980)\pi^0$ & $-34.0^{+3.5+3.1+65.5}_{-3.9-2.9-47.2}$  &  \non \\
 $B^-\to a_0^0(980)K^-$ & $-13.3^{+2.7+7.3+56.8}_{-2.7-6.1-40.1}$ &  & $\ov B^0\to a_0^+(980)K^-$ &
 $-19.1^{+5.3+10.1+75.2}_{-6.4-~8.6-41.2}$ &   \non \\
 $B^-\to a_0^-(980)\ov K^0$ & $0.63^{+0.74+54.22+2.49}_{-0.57-~0.49-2.86}$ &  & $\ov B^0\to a_0^0(980)\ov K^0$
 & $-7.6^{+1.9+58.2+35.6}_{-3.9-~3.0-30.5}$ &  \non \\
 $B^-\to a_0^0(980)\pi^-$ & $-8.8^{+0.6+0.0+0.9}_{-0.6-0.0-0.8}$ &  & $\ov B^0\to a_0^+(980)\pi^-$
 & $1.1^{+0.1+0.1+16.5}_{-0.1-0.0-16.9}$ &    \non \\
 $B^-\to a_0^-(980)\pi^0$ & $-32.8^{+5.6+3.2+20.3}_{-6.2-2.8-21.6}$ &  & $\ov B^0\to a_0^-(980)\pi^+$
 & $-5.0^{+1.0+0.6+16.6}_{-0.9-0.5-15.5}$ &  \non \\
 & & &  $\ov B^0\to a_0^0(980)\pi^0$ &
 $-30.9^{+4.6+1.9+30.5}_{-5.6-1.8-24.7}$ & \non \\
 $B^-\to a_0^0(1450)K^-$ & $-1.79^{+0.69+0.64+16.61}_{-0.82-0.06-17.75}$ &  & $\ov B^0\to a_0^+(1450)K^-$ &
 $-0.65^{+0.41+0.68+11.29}_{-0.54-0.23-23.80}$ &  \non \\
 $B^-\to a_0^-(1450)\ov K^0$ & $0.22^{+0.04+0.06+~0.59}_{-0.03-0.03-47.94}$ &  & $\ov B^0\to a_0^0(1450)\ov K^0$
 & $-0.74^{+0.32+0.23+~8.2}_{-0.39-0.05-20.7}$ &  \non \\
 $B^-\to a_0^0(1450)\pi^-$ & $-13.1^{+0.9+0.0+4.9}_{-0.9-0.0-4.3}$ & & $\ov B^0\to a_0^+(1450)\pi^-$
 & $-0.21^{+0.37+0.18+82.8}_{-0.46-0.28-83.6}$ &  \non \\
 $B^-\to a_0^-(1450)\pi^0$ & $-24.6^{+3.4+2.4+25.3}_{-3.8-2.1-23.5}$ &  & $\ov B^0\to a_0^-(1450)\pi^+$
 & $-5.2^{+3.4+2.3+62.0}_{-2.9-1.7-66.9}$ &  \non \\
 & & &  $\ov B^0\to a_0^0(1450)\pi^0$ &
 $-13.8^{+3.1+2.1+35.7}_{-3.8-1.9-20.0}$ & \non \\
 $B^-\to \ov K^{*0}_0(1430)\pi^-$ & $1.3^{+0.1+0.0+5.9}_{-0.1-0.0-4.8}$ & $5.5^{+3.4}_{-2.2}$ &
 $\ov B^0\to K^{*-}_0(1430)\pi^+$  & $0.21^{+0.06+0.05+7.05}_{-0.06-0.04-7.21}$ & $7\pm14$ \non \\
 $B^-\to K^{*-}_0(1430)\pi^0$ & $3.0^{+0.4+0.5+10.7}_{-0.4-0.4-~7.8}$ &  &
 $\ov B^0\to \ov K^{*0}_0(1430)\pi^0$  & $-1.9^{+0.4+0.4+12.0}_{-0.5-0.5-14.9}$ &
 $-15\pm11$  \non \\
 $B^-\to K^{*-}_0(1430)\eta$ & $1.5^{+0.1+0.2+1.7}_{-0.1-0.2-1.9}$ & $-19\pm17$ &
 $\ov B^0\to \ov K^{*0}_0(1430)\eta$  & $2.0^{+0.1+0.2+2.3}_{-0.1-0.2-1.3}$ & $6\pm13$ \non \\
 $B^-\to K^{*-}_0(1430)\eta'$ & $1.2^{+0.3+0.1+8.7}_{-0.2-0.3-6.3}$ & $6\pm20$ &
 $\ov B^0\to \ov K^{*0}_0(1430)\eta'$  & $1.2^{+0.1+0.1+17.9}_{-0.1-0.4-13.5}$ & $-19\pm17$
\end{tabular}
}
\footnotetext[1]{This is the naive (uncorrelated) average of the direct \CP asymmetries $-0.29\pm0.20$ obtained from $B^0\to K^+K^-K_S$ and $-0.07\pm0.14$ from $B^0\to \pi^+\pi^- K_S$ (see Table \ref{tab:exptCP}).}
\end{ruledtabular}
\end{table}

\begin{table}[t]
\caption{Same as Table \ref{tab:asySV} except for $B$ decays to
a scalar meson and a vector meson.}
\label{tab:asySV}
\begin{ruledtabular}
\footnotesize{
\begin{tabular}{l r c |l  r c}
Mode & Theory & Expt & Mode & Theory & Expt  \\
\hline
 $B^-\to f_0(980)K^{*-}$ & $-3.3^{+0.2+0.7+5.7}_{-0.2-0.8-0.9}$ & $-15\pm12$ & $\ov B^0\to f_0(980)\ov K^{*0}$ & $3.6^{+0.3+0.1+0.7}_{-0.2-0.1-1.3}$  & $7\pm10$ \non \\
 $B^-\to f_0(980)\rho^-$ & $73.0^{+3.2+5.1+5.9}_{-3.4-7.0-6.1}$ &  &
 $\ov B^0\to f_0(980)\rho^0$ & $68.2^{+2.4+0.6+~22.5}_{-2.3-0.5-138.3}$  &   \non \\
 & & &  $\ov B^0\to f_0(980)\omega$ & $55.0^{+1.6+0.5+~28.5}_{-1.5-0.4-108.4}$  &   \non \\
 $B^-\to a_0^0(980)K^{*-}$ & $4.6^{+0.6+0.3+~2.0}_{-0.6-0.3-18.2}$ &  & $\ov B^0\to a_0^+(980)K^{*-}$ &
 $3.8^{+0.6+0.3+~0.1}_{-0.6-0.3-21.6}$ &   \non \\
 $B^-\to a_0^-(980)\ov K^{*0}$ & $0.4^{+0.0+0.0+0.0}_{-0.0-0.0-0.3}$ &  & $\ov B^0\to a_0^0(980)\ov K^{*0}$
 & $0.1^{+64.1+72.7+~2.0}_{-79.3-73.3-18.2}$ &  \non \\
 $B^-\to a_0^0(980)\rho^-$ & $-10.7^{+0.6+1.1+19.0}_{-0.7-1.4-~1.4}$ &  & $\ov B^0\to a_0^+(980)\rho^-$
 & $-1.3^{+0.1+0.0+8.2}_{-0.1-0.0-0.0}$ &   \non \\
 $B^-\to a_0^-(980)\rho^0$ & $4.2^{+2.7+1.8+28.1}_{-2.1-1.8-12.4}$ &  & $\ov B^0\to a_0^-(980)\rho^+$
 & $-12.7^{+0.7+0.0+~6.3}_{-0.6-0.0-24.5}$ &  \non \\
 $B^-\to a_0^-(980)\omega$ & $6.2^{+2.9+0.6+47.5}_{-2.1-1.0-~7.1}$ & & $\ov B^0\to a_0^0(980)\rho^0$ &
 $7.8^{+0.7+0.2+12.5}_{-0.6-0.1-11.7}$ & \non \\
 & & &  $\ov B^0\to a_0^0(980)\omega$ &
 $-72.8^{+8.9+0.3+23.4}_{-6.4-0.3-22.0}$ & \non \\
 $B^-\to a_0^0(1450)K^{*-}$ & $2.6^{+0.6+0.3+~3.6}_{-0.5-0.4-20.2}$ &  & $\ov B^0\to a_0^+(1450)K^{*-}$ &
 $42.6^{+2.3+4.0+~~0.6}_{-2.5-5.1-140.4}$ &  \non \\
 $B^-\to a_0^-(1450)\ov K^{*0}$ & $0.36^{+0.02+0.02+0.08}_{-0.02-0.02-0.88}$ &  & $\ov B^0\to a_0^0(1450)\ov K^{*0}$
 & $1.0^{+0.2+0.0+3.3}_{-0.2-0.1-9.6}$ &  \non \\
 $B^-\to a_0^0(1450)\rho^-$ & $-19.1^{+1.9+2.7+45.2}_{-2.3-3.8-~5.4}$ & & $\ov B^0\to a_0^+(1450)\rho^-$
 & $-1.4^{+0.1+0.0+19.0}_{-0.1-0.0-~0.4}$ &  \non \\
 $B^-\to a_0^-(1450)\rho^0$ & $-1.6^{+1.5+1.3+37.2}_{-1.2-1.3-18.4}$ &  & $\ov B^0\to a_0^-(1450)\rho^+$
 & $-13.3^{+0.9+0.0+~7.1}_{-0.8-0.0-22.9}$ &  \non \\
 $B^-\to a_0^-(1450)\omega$ & $1.6^{+2.3+1.5+54.7}_{-1.7-1.7-18.1}$ & &  $\ov B^0\to a_0^0(1450)\rho^0$ &
 $6.4^{+0.7+0.3+12.2}_{-0.6-0.3-~5.3}$ & \non \\
 & & &  $\ov B^0\to a_0^0(1450)\omega$ &
 $-31.9^{+2.5+0.8+30.8}_{-2.9-1.0-21.3}$ & \non \\
 $B^-\to K^{*-}_0(1430)\phi$ &  $0.64^{+0.02+0.01+0.32}_{-0.03-0.01-1.49}$ & $4\pm15$
 & $\ov B^0\to \ov K^{*0}_0(1430)\phi$ &
 $0.43^{+0.04+0.01+~3.61}_{-0.04-0.00-20.2}$ & $20\pm15$ \non \\
 $B^-\to \ov K^{*0}_0(1430)\rho^-$ & $0.32^{+0.34+0.33+0.71}_{-0.29-0.31+1.09}$ & $$ &
 $\ov B^0\to K^{*-}_0(1430)\rho^+$  & $1.1^{+0.0+0.2+0.2}_{-0.0-0.2-1.2}$ & $$ \non \\
 $B^-\to K^{*-}_0(1430)\rho^0$ & $1.6^{+0.6+0.1+2.8}_{-0.6-0.1-6.2}$ &  &
 $\ov B^0\to \ov K^{*0}_0(1430)\rho^0$  & $0.54^{+0.45+0.02+3.76}_{-0.46-0.02-1.80}$ &
 $$  \non \\
 $B^-\to K^{*-}_0(1430)\omega$ & $0.55^{+0.35+0.08+0.32}_{-0.34-0.08-2.49}$ & $-10\pm9$ &
 $\ov B^0\to \ov K^{*0}_0(1430)\omega$  & $0.03^{+0.37+0.01+0.29}_{-0.35-0.01-3.00}$ & $-7\pm9$  \non \\
\end{tabular}
}
\end{ruledtabular}
\end{table}

\section{Results and Discussions}

The calculated branching fractions and \CP asymmetries of $B\to SP$ and $B\to SV$ decays are summarized in Tables \ref{tab:BRSP}--\ref{tab:asySV}. The defaulted values of the  parameters $\rho_{A,H}$ and $\phi_{A,H}$ introduced in Eq. (\ref{eq:XA}) are set to zero; that is, the central values (or ``default" results) correspond to $\rho_{A,H}=0$ and $\phi_{A,H}=0$.
The first theoretical error shown in the Tables for QCDF results is due to the
variation of $B_{1,3}$ and $f_S$, the second error comes from the
uncertainties of form factors and the strange quark mass,
while the third error from the power corrections due to weak
annihilation and hard spectator interactions.

In order to compare theory
with experiment for decays involving $f_0(980)$ or $a_0(980)$ and $a_0(1450)$, we need an input for $\B(f_0(980)\to \pi^+\pi^-)$ or $\B(a_0\to\pi\eta)$. To do this for $f_0(980)$, we shall use
the BES measurement \cite{BES}
 \be
 {\Gamma(f_0(980)\to \pi\pi)\over \Gamma(f_0(980)\to \pi\pi)+\Gamma(f_0(980)\to K\ov
K)}=0.75^{+0.11}_{-0.13}\,.
 \en
Assuming the dominance of the $f_0(980)$ width by $\pi\pi$
and $K\ov K$ and applying isospin relation, we obtain \footnote{This is in agreement with the value of $\B(f_0(980)\to \pi^+\pi^-)=0.46\pm0.06$ obtained in \cite{LHCb:theta}.}
 \be
 \B(f_0(980)\to \pi^+\pi^-)=0.50^{+0.07}_{-0.09}\,, \qquad \B(f_0(980)\to
 K^+K^-)=0.125^{+0.018}_{-0.022}\,.
 \en

For $a_0(980)$, we shall apply the Particle Data Group (PDG) average
$\Gamma(a_0\to K\ov K)/\Gamma(a_0\to \pi\eta)=0.183\pm 0.024$
\cite{PDG} to obtain
\be
 \B(a_0(980)\to \eta\pi)=0.845\pm0.017\,.
\en
For $a_0(1450)$, we use $\Gamma(a_0(1450)\to \pi\eta')/\Gamma(a_0(1450)\to \pi\eta)=0.35\pm0.16$ and $\Gamma(a_0(1450)\to K\bar K)/\Gamma(a_0(1450)\to\pi\eta)=0.88\pm0.23$ to obtain
\be
1/\B(a_0(1450)\to\pi\eta)=1.52\pm0.13\,.
\en

\subsection{Decays involving a $K_0^*(1430)$ state}
Among the hadronic 2-body $B$ decays with a $K_0^*(1430)$ in the final state, only $B\to K_0^*\eta^{(')}$ and $B\to K_0^*\phi$ are sensitive to the $B\to K_0^*(1430)$ transition form factors $F_0^{BK_0^*}$ and $F_1^{BK_0^*}$, respectively. It turns out that the measurement of $B\to K_0^*\eta^{(')}$ favors a smaller $F_0^{BK_0^*}(0)$, while the data of $B\to K_0^*\phi$ prefer a smaller $F_1^{BK_0^*}(0)$ inferred from the neutral mode and a slightly larger $F_1^{BK_0^*}(0)$ from the charged mode. Since $F_1(0)=F_0(0)$, this means that it is preferable to fit the data of $B\to K_0^*\eta^{(')}$ and $B\to K_0^*\phi$ by a smaller $B\to K_0^*(1430)$ transition form factor.
In this work we shall use $F_0^{BK_0^*}(0)\sim 0.26$ obtained in the covariant light-front quark model \cite{CCH}.

\subsubsection{$B\to K_0^*(1430)\eta^{(')}$ decays}
Following \cite{CC:Keta}, we write
\be
A_{B\to K_0^*\eta} &=& X^{(BK_0^*,\eta_q)} C_1+X^{(BK_0^*,\eta_s)}C_2+X^{(B\eta_q,K_0^*)}C_3, \non \\
A_{B\to K_0^*\eta'} &=& X^{(BK_0^*,\eta'_q)}C_1+X^{(BK_0^*,\eta'_s)}C_2+X^{(B\eta'_q,K_0^*)}C_3,
\en
where $C_1$, $C_2$ and $C_3$ terms correspond to Figs. 1(a), 1(b) and 1(c), respectively, and $X^{(BS,P)}$, $X^{(BP,S)}$ are factorizable terms defined in Eq. (\ref{eq:X}).
The expressions of $C_i$'s in terms of the parameters $\alpha_i^p$ can be found in \cite{CC:Keta}. Because of the small vector decay constant of $K_0^*(1430)$, $X^{(B\eta^{(')}_q,K_0^*)}$ is suppressed relative to
$X^{(BK_0^*,\eta^{(')}_q)}$ and $X^{(BK_0^*,\eta^{(')}_s)}$. However, the $C_3$ term gains a large enhancement from $\alpha_4^c(\eta_q K_0^*)$ due to the fact that the chiral factor $r_\chi^{K^*_0}= 12.3$ at $\mu=4.2$ GeV is larger than $r_\chi^K=1.5$ by one order of magnitude
owing to the large mass of $K_0^*(1430)$. It follows that $\alpha_4^c(\eta_q K_0^*)$ is much greater than $\alpha_4^c(K_0^*\eta_s)$ and $\alpha_4^c(K_0^*\eta_q)$. As a result, the amplitude of Fig. 1(c) is comparable to that of Fig. 1(a).

\begin{figure}[t]
\begin{center}
\includegraphics[width=0.80\textwidth]{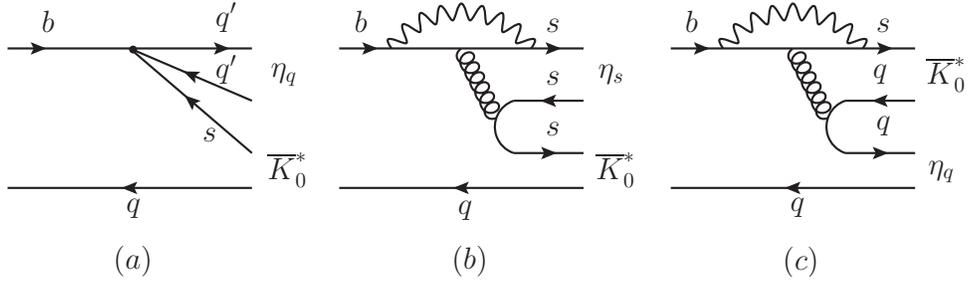}
\vspace{0.0cm}
\caption{Three different penguin contributions to $\ov B\to \ov K_0^*(1430)\eta^{(')}$. Fig. 1(a) is induced by the penguin operators $O_{3,5,7,9}$.
} \label{fig:Keta} \end{center}
\end{figure}


Because of the large magnitude of  $\alpha_3(K_0^*\eta_{q,s})$ and the large cancellation between $\alpha_3(K_0^*\eta_s)$ and $\alpha_4(K_0^*\eta_s)$ in $C_2$, $B\to K_0^*\eta^{(')}$ decays are dominated by the contributions from Figs. 1(a) and 1(c) \cite{CC:Keta}.
Therefore, the penguin diagrams Figs. 1(a) and 1(c) contribute constructively to both $K_0^*\eta$ and $K^*_0\eta'$ with comparable magnitudes. Since $X^{(BK_0^*,\eta_q)}/X^{(BK_0^*,\eta'_q)}=X^{(B\eta_q, K_0^*)}/X^{(B\eta'_q, K_0^*)}=\cot\phi\approx 1.23$, it is clear that $A_{B\to K_0^*\eta}/A_{B\to K_0^*\eta'}\approx \cot\!\phi$ and hence $B\to K_0^*\eta$ should have a rate larger than $B\to K_0^*\eta'$ in scenario 2 as the mixing angle $\phi$ is less than $45^\circ$.

As mentioned in the passing, we use the form factor $F_0^{BK_0^*}(0)\sim 0.26$ in this work. If a large form factor, say, $F_0^{BK_0^*}(0)\sim 0.45$ \cite{Han}, is employed, we will have $\B(B\to K_0^*(1430)\eta)\sim 21\times 10^{-6}\gg \B(B\to K_0^*(1430)\eta)\sim 3.2\times 10^{-6}$. This indicates that a small form factor for $B\to K_0^*(1430)$ transition is preferable in this case.

A recent pQCD calculation \cite{Xiao:K0steta} shows $\B(B\to K_0^*\eta')\approx 75\times 10^{-6}\gg\B(B\to K_0^*\eta)$, in sharp contrast to the experimental measurements  $\B(B\to K_0^*\eta')<\B(B\to K_0^*\eta)$ and $\B(B\to K_0^*\eta')\sim 6\times 10^{-6}$ (see Table \ref{tab:BRSP}).

\subsubsection{$B\to K_0^*(1430)\pi$ decays}
It is clear from Table \ref{tab:BRSP} that while the predicted branching fraction of $\ov B^0\to \ov K_0^{*0}(1430)\pi^0$ is consistent with the data, the calculated rates of $\ov K_0^{*0}(1430)\pi^-$ and $K_0^{*-}(1430)\pi^+$ are too small compared to experiment. Under the isospin limit, it is naively expected that
\be \label{eq:ratios}
 && R_1\equiv {\B(\ov B^0\to \ov K_0^{*0}(1430)\pi^0)\over \B(\ov B^0\to
 K_0^{*-}(1430)\pi^+)} = {1\over 2}, \qquad
 R_2\equiv {\B(B^-\to K_0^{*-}(1430)\pi^0)\over \B(B^-\to \ov
 K_0^{*0}(1430)\pi^-)} ={1\over 2}, \non \\
 && \qquad\qquad\qquad  R_3\equiv {\tau(B^0)\B(B^-\to \ov K_0^{*0}(1430)\pi^-)\over \tau(B^-)\B(\ov B^0\to K_0^{*-}(1430)\pi^+)} = 1\,.
\en
However, the first two relations are not borne out by experiment. In principle, the data of $K_0^{*-}(1430)\pi^+$ and $\ov K_0^{*0}(1430)\pi^-$ can be accommodated by taking into account the
power corrections due to the non-vanishing $\rho_A$ and $\rho_H$
from weak annihilation and hard spectator interactions,
respectively. However, this will affect the agreement between theory and experiment for $B\to K_0^*(1430)\eta^{(')}$ and $B\to K_0^*(1430)\rho$. Indeed, a global fit of $\rho_A$ and $\phi_A$ to the $B\to SP$ data shown in Table \ref{tab:BRSP} yields $\rho_A=0.15$ and $\phi_A=82^\circ$ with $\chi^2=8.3$\,. The best fitted results are $\B(B^-\to \ov K_0^{*0}(1430)\pi^-)\sim 13.4\times 10^{-6}$ and $\B(\ov B^0\to  K_0^{*-}(1430)\pi^+)\sim 14.1\times 10^{-6}$ which are very close to the QCDF predictions with $\rho_A=0$ and $\phi_A=0$.

The measured branching fractions of $\ov B^0\to K_0^{*-}(1430)\pi^+$ and $B^-\to \ov K_0^{*0}(1430)\pi^-$ are of order $30\times 10^{-6}$ and $50\times 10^{-6}$ by BaBar and Belle, respectively (see Table \ref{tab:exptBR}), though they are consistent with each other within errors. This is probably ascribed to the fact that the definitions of the $K_0^*(1430)$ and nonresonant contributions by BaBar and Belle are different.
While Belle employed the Brei-Wigner parametrization to describe the $K_0^*(1430)$ resonance, BaBar used the LASS parametrization  to describe the $K\pi$ $S$-wave and
the nonresonant component by a single amplitude suggested by the
LASS collaboration \cite{LASS} to describe the scalar amplitude in elastic
$K\pi$ scattering. Since the LASS parametrization is valid up to the $K\pi$ invariant mass of order 1.8 GeV, BaBar introduced a phase-space nonresonant component to describe an excess of signal events at higher $K\pi$ invariant mass. Hence, the BaBar definition for the $K^*_0(1430)$ includes an effective range term to account for the low
$K\pi$ $S$-wave while for the Belle parametrization, this component is absorbed into the nonresonant piece.
In order to compare the BaBar results with the Belle ones determined from the Breit-Wigner parametrization,
it would be more appropriate to consider the Breit-Wigner component only of the LASS parametrization. Indeed, the BaBar results for $B\to K_0^*(1430)\pi$ quoted in Table \ref{tab:BRSP} are obtained from $(K\pi)_0^{*0}\pi^0$ and $(K\pi)_0^{*-}\pi^+$ by subtracting the elastic range term from the $K\pi$ $S$-wave \cite{BABAR:Kppimpi0,BaBar:KSpippim}. However, the discrepancy between BaBar and Belle for the $K_0^*\pi$ modes still remains and it is crucial to resolve this important issue.

Contrary to the decay mode $\ov B^0\to K_0^{*-}(1430)\pi^+$ whose rate is predicted too small compared to the data, the calculated $\B(\ov B^0\to K_0^{*-}(1430)\rho^+)$ is in accordance with experiment. Therefore, it is a puzzle why the QCDF approach works well for $K_0^{*-}(1430)\rho^+$ but not for $K_0^{*-}(1430)\pi^+$, whereas it is the other way around for pQCD.

It appears that the calculated branching fractions of $\ov K_0^{*0}\pi^-$ and $K_0^{*-}\pi^+$ by pQCD \cite{Xiao:K0steta} are in better agreement with the data (see Table \ref{tab:BRSP}). However, the predicted rates of $B\to K_0^*\rho$ in this approach are too small as we shall see shortly below.

\subsubsection{$B\to K_0^*(1430)(\rho,\omega,\phi)$ decays}

We see from Table \ref{tab:BRSV} that the calculated $B\to K_0^*(1430)(\phi,\rho,\omega)$ rates are in good agreement with experiment, though the central value of $B^-\to K_0^{*-}(1430)\phi$ is smaller than the data.
It is obvious that the data of $B\to K_0^*(1430)(\phi,\rho,\omega)$ can be well accommodated without introducing penguin annihilation effects characterized by the parameters $\rho_A$ and $\phi_A$ to the central values.

Note that the predicted branching fractions of $B\to K_0^*(1430)\phi$ and $B\to K_0^*(1430)\rho$ in this work are substantially smaller than those shown in Table IV of \cite{CCY:SV}. We found that the rate of $B\to K_0^*(1430)\phi$ is sensitive to the scale of $\mu$ and is large at $\mu=m_b/2$ which is the scale used in \cite{CCY:SV}. While $\B(B\to K_0^*(1430)\rho)$ is stable against the choice of $\mu$, we found a sign mistake in the computer code of the previous work \cite{CCY:SV}; the sign in front of $\Phi_{M_1}(\eta)$ in Eqs. (3.7) and (3.8) of \cite{CCY:SV} should read $\mp$ instead of $\pm$. As a consequence, the calculated $\B(B\to K_0^*(1430)\rho)$ in \cite{CCY:SV} were too large.

As stated before, we use $F_0^{BK_0^*}(0)\sim 0.26$  in order to account for the data of $B\to K_0^*(1430)\eta^{(')}$. The predicted $\ov B^0\to\ov K^{*0}_0(1430)\phi$ is in good agreement with experiment, while the central value of $\B(B^-\to K^{*-}_0(1430)\phi)$ is smaller than experiment. Nevertheless, we notice that the current data imply a large disparity between $K^{*0}_0(1430)\phi$ and $K^{*-}_0(1430)\phi$.

For comparison, we see from Table \ref{tab:BRSV} that in the pQCD approach \cite{Zhang:K0strho,CSKim}, the predicted branching fractions are too small for $B\to K_0^*(1430)(\rho,\omega)$ and too large for $B\to K_0^*(1430)\phi$  compared to experiment. It is interesting to see that the rates of $B\to K_0^*\pi$ are larger than that of $B\to K_0^*\rho$ within the pQCD framework, whereas it is the other way around in QCDF.

\subsection{Decays involving a $a_0$ meson}

As stated before, we shall work in scenario 1 for the light scalar meson $a_0(980)$  and scenario 2 for the heavy one $a_0(1450)$ so that both of them are ground $q\bar q$ bound states. We see from Table \ref{tab:BRSP} that in general $\B(B\to a_0(980)K)$ is only of order $10^{-7}$. \footnote{In \cite{CCY:SP}, the predicted rates are too large for $a_0(980)K$ in scenario 1 and too small for $a_0(1450)K$ in scenario 2. This is ascribed to a typo appearing in the computer code for the chiral factor $r_\chi^K$.}
This may explain why $a_0(980)$ has not been seen thus far in hadronic $B$ decays, whereas plenty of $a_0(980)$ events have been observed in $D$ decays. Notice that $\B(B\to a_0(980)K)$ are predicted to fall into the range of $(4\sim 10)\times 10^{-6}$ in the pQCD approach \cite{Shen} and all of them are ruled out by experiment except $\ov B\to a_0^0\ov K^0$.

Contrary to $B\to PP$ decays where the production of $\pi^+K^-$ is substantially greater than $\pi^+\pi^-$, it is expected in QCDF that $a_0(980)^+K^-$ has a rate much smaller than $a_0(980)^+\pi^-$. Consider the interference between the QCD penguin amplitude governed by  $a_4^p(a_0P)-r_\chi^P a_6^p(a_0P)$ and the penguin annihilation amplitude proportional to $(V_{ub}V_{uq}^*+V_{cb}V_{cq}^*)f_Bf_{a_0}f_Pb_3(a_0P)$ for $P=K,\pi$ with $q=d,s$. Since $a_{4,6}$, $b_3$ and $f_{a_0^+}$ are all negative, $\sum_{p}\lambda_p^{(s)}=0.04$ and $\sum_{p}\lambda_p^{(d)}=-0.008-0.003i$,  the interference is destructive for $\ov B^0\to a_0^+K^-$ and constructive for $\ov B^0\to a_0^+\pi^-$. This explains why the former has a rate much smaller than the latter in QCDF. By contrast, pQCD  predicts the other way around \cite{Zhang:a0pi,Shen}.

Since the decay amplitudes of the tree-dominated decays $\bar B^0\to a_0^+\pi^-$ and $\bar B^0\to a_0^-\pi^+$ are proportional to $a_1 f_\pi F_0^{Ba_0}$ and $a_1 f_{a_0^-}F_0^{B\pi}$, respectively, it is thus expected that $\B(\bar B^0\to a_0^+\pi^-)\gg \B(\bar B^0\to a_0^-\pi^+)$ for both $a_0=a_0(980)$ and $a_0(1450)$ as the decay constant $f_{a_0^-}$ is very small. Moreover, the mode $a_0^+\pi^-$ should have the largest rate among various $\ov B\to a_0\pi$ decays. Therefore, the pQCD results for $a_0^0(980)\pi^-$ and $a_0^+(980)\pi^-$ \cite{Zhang:a0pi} are at odd with the expectation: The rate of $a_0^0(980)\pi^-$ is predicted to be larger than $a_0^+(980)\pi^-$ and $a_0^-(980)\pi^+$ is not very suppressed relative to $a_0^+(980)\pi^-$.

The decay rate of $\ov B^0\to a_0^\pm(980)\pi^\mp$ is not simply the sum of $\Gamma(\ov B^0\to a_0^+(980)\pi^-)$ and $\Gamma(\ov B^0\to a_0^-(980)\pi^+)$ because of the interference of $a_0^+(\to \eta\pi^+)\pi^-$ with $a_0^-(\to \eta\pi^-)\pi^+$.
In principle, one should study the 3-body decay $\ov B^0\to\eta\pi^+\pi^-$ and then apply the narrow-width approximation
\be
\B(\ov B^0\to\eta\pi^+\pi^-)=\B(\ov B^0\to a_0^\pm \pi^\mp)\B(a_0^\pm\to\eta\pi^\pm)
\en
to determine the branching fraction of $\ov B^0\to a_0^\pm \pi^\mp$. While a direct evaluation of the decay amplitude of $\ov B^0\to\eta\pi^+\pi^-$ can be done in the factorization approach (see e.g. \cite{CCSfsi}), we will simplify the calculation by assuming that the amplitude $A(\ov B^0\to a_0^\pm \pi^\mp)$ is the sum of $A(\ov B^0\to a_0^+\pi^-)$ and $A(\ov B^0\to a_0^- \pi^+)$.
Since the decay constants $f_{a_0^+}$ and $f_{a_0^-}$ are of opposite sign [see Eq. (\ref{eq:a0decaycon})], the interference between $a_0^+\pi^-$ and $a_0^-\pi^+$ is destructive and we find $\B(\ov B^0\to a_0(980)^\pm\pi^\mp)= (4.7^{+2.3}_{-1.3})\times 10^{-6}$ in QCDF. Within theoretical uncertainties this is consistent with the current experimental limit $3.7\times 10^{-6}$ set by BaBar \cite{BaBar:a0pi}.
Likewise, the calculated  $\B(\ov B^0\to a_0(1450)^\pm\pi^\mp)\sim 1.3\times 10^{-6}$ is also in accordance with the limit $3.5\times 10^{-6}$ \cite{BaBar:a0pi}.

So far the results of $\B(B\to a_0(980)\pi)$ and $\B(B\to a_0(980)K)$  in QCDF are all consistent with the experimental limits. Hence, at this moment, we cannot conclude on the 2-quark or 4-quark nature of $a_0(980)$.
Nevertheless, as stressed in \cite{CCY:SP}, if the measured rate of $a_0^\pm(980)\pi^\mp$ is at
the level of $1\times 10^{-6}$ or even smaller, this will
imply a substantially smaller $B\to a_0(980)$ form factor than the
$B\to\pi$ one. In this case, the four-quark explanation of the $a_0(980)$ will be preferred to account for the $B\to a_0(980)$
form factor suppression. Since $a_0(1450)$
can be described by the $q\bar q$ quark model, the study of
$a_0^\pm(1450)\pi^\mp$ relative to $a_0^\pm(980)\pi^\mp$ can provide a
more strong test on the quark content of $a_0(980)$. We see that if the
branching fractions of both $\ov B^0\to a_0^\pm(1450)\pi^\mp$ and $\ov B^0\to a_0^\pm(980)\pi^\mp$ are measured at
the level of $1\times 10^{-6}$, it
will be likely to imply a 2-quark nature for $a_0(1450)$ and a
four-quark assignment for $a_0(980)$.

Among the decays $\ov B\to a_0(980)\rho$ and $\ov B\to a_0(1450)\rho$, it is clear that $a_0^+(980)\rho^-$ and $a_0^+(1450)\rho^-$ have the largest rates.
Since the branching fraction of $\ov B^0\to a_0^\pm(980)\rho^\mp$ is predicted to be large in QCDF, of order $23\times 10^{-6}$, a measurement of this mode may give the first observation of the $a_0(980)$ production in $B$ decays.
Notice that the calculated rates of $B^-\to a_0^0(1450)\rho^-$ and $\ov B^0\to a_0^+(1450)\rho^-$ in the pQCD approach \cite{Zhang:a01450rho} are unreasonably too large.

\subsection{Scenario 1 for $a_0(1450)$ and $K_0^*(1430)$}

In Table \ref{tab:theoryBRS1} we show the branching fractions of $B\to SP,SV$ decays with $S=a_0(1450),K_0^*(1430)$ in scenario 1 where scalar mesons $a_0(1450)$ and $K_0^*(1430)$ are treated as the first excited states of low lying light $q\bar q$ scalars $a_0(980)$ and $K_0^*(800)$. It is evident that this scenario for heavy scalar mesons is ruled out by experiment. \footnote{An exception is the decay $B\to K_0^*(1430)\eta'$: the calculated rates in scenario 1 are consistent with the data.}
For example, the predicted $\B(\ov B^0\to a_0^\pm(1450)\pi^\mp)$ is too large and the branching fractions of $B\to K_0^*(1430)(\rho,\omega)$  are too small compared to the data. In Tables \ref{tab:BRSP} and \ref{tab:BRSV} we have found a better agreement between theory and experiment for  scalar mesons above 1 GeV in scenario 2 in which $a_0(1450)$ and $K_0^*(1450)$ are lowest lying  $q\bar q$ states. This also implies that the light scalars $K_0^*(800),a_0(980)$ and $f_0(980)$ are preferred to be four-quark bound states.

\subsection{Decays involving a $f_0(980)$ meson}

\begin{figure}[t]
\begin{center}

\centerline{\epsfig{figure=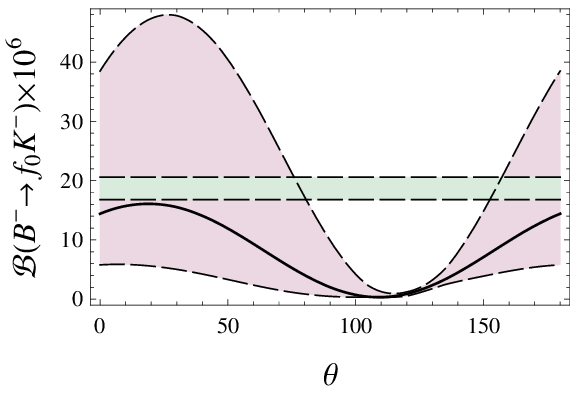,width=7.0cm}
              \hspace{0.7cm}
              \epsfig{figure=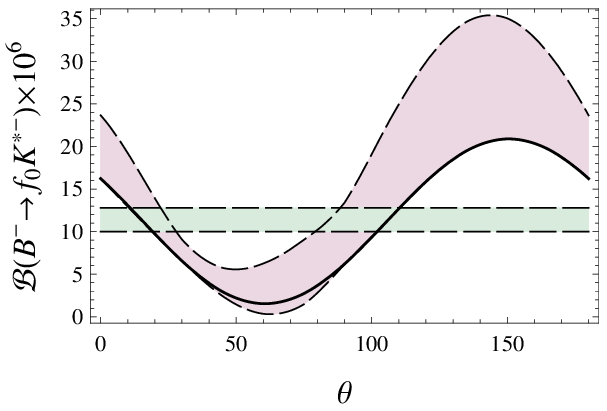,width=7.0cm}
              }
\centerline{(a)
              \hspace{7.0cm}
              (b)
              }
\vspace{0.0cm}
\caption{Branching fractions of (a) $B^-\to f_0(980)K^-$ and (b)
    $B^-\to f_0(980)K^{*-}$  versus the
    mixing angle $\theta$ of strange and nonstrange components of $f_0(980)$,
    where the middle bold solid curve inside the allowed region
    corresponds to the central value. Theoretical errors due to the
    penguin annihilation are taken into account.
    The horizontal band within the dashed lines
    shows the experimentally allowed region with one sigma error.
} \label{fig:f0K} \end{center}
\end{figure}

The penguin-dominated $B\to f_0(980)K^{(*)}$ decays receive three
distinct types of factorizable contributions: one from the $K^{(*)}$
emission, one from the $f_0$ emission with the $s\bar s$ content,
and the other from the $f_0$ emission with the $n\bar n$
component; see Eq. (\ref{eq:SDAmp}). Therefore, $\B(B\to f_0(980)K^{(*)})$ depends on the mixing angle
$\theta$ of strange and nonstrange components of the $f_0(980)$.
In Fig. \ref{fig:f0K}
branching fractions of $B^-\to f_0(980)K^-$ and
$B^-\to f_0(980)K^{*-}$  are plotted  versus the $f_0(980)-f_0(500)$ mixing angle $\theta$ defined in Eq. (\ref{eq:f0mixing}). We see that both $f_0 K^{-}$ and $f_0 K^{*-}$ rates can be accommodated with $\theta$ in the vicinity of  $20^\circ$ without introducing $1/m_b$ power corrections from penguin annihilation. This range of the mixing angle is consistent with the limit $|\theta|<30^\circ$ set recently by LHCb.

\begin{figure}[]
\begin{center}

\centerline{\epsfig{figure=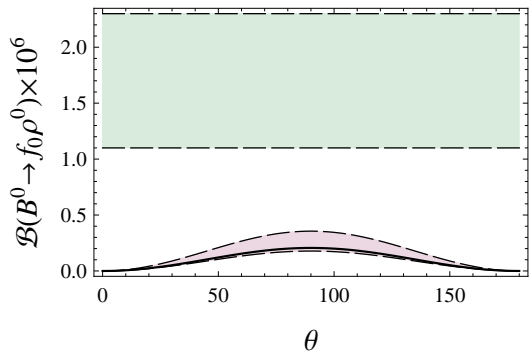,width=7.0cm}
              }
\vspace{0.0cm}
\caption{Same as Fig. \ref{fig:f0K} except for the decay $B^0\to f_0(980)\rho^0$.
} \label{fig:f0Krho0} \end{center}
\end{figure}

For definiteness, we show in Tables \ref{tab:BRSP}-\ref{tab:asySV} the branching fractions and \CP asymmetries of $B\to f_0(980)K^{(*)}$ for $\theta=17^\circ$. With this mixing angle, the calculated branching fraction of $\ov B^0\to f_0(980)\rho^0$ by QCDF or pQCD is much smaller than the Belle measurement $\B(B\to f_0(980)\rho^0)=(1.7\pm0.6)\times 10^{-6}$ \cite{Belle:f0rho0} as shown in Fig. \ref{fig:f0Krho0}.  Since the BaBar measurement yielded only an upper bound $\B(B\to f_0(980)\rho^0)<0.8\times 10^{-6}$ \cite{BaBar:f0rho0}, the experimental issue remains to be resolved.

In order to make quantitative calculations
for $B\to f_0(980)K^{(*)}$, we have assumed the conventional 2-quark
description of the light scalar mesons. However, the fact that their rates can be accommodated in the 2-quark
picture for $f_0(980)$ does not mean that the measurements of $B\to
f_0K^{(*)}$ can be used to distinguish between the 2-quark and 4-quark
assignment for $f_0(980)$. As discussed in \cite{CCY:SP,Brito},
the number of the
quark diagrams for the penguin contributions to $B\to f_0(980)K^{(*)}$
in the four-quark scheme for $f_0(980)$ is
two times as many as that in the usual 2-quark picture.  Therefore, there is no reason that the $B\to f_0(980)K^{(*)}$ rate will
be suppressed if $f_0$ is a four-quark state. However, in
practice, it is difficult to give quantitative predictions based
on this scenario as the nonfactorizable diagrams are usually not
amenable. Moreover, even for the factorizable contributions, the
calculation of the $f_0(980)$ decay constant and its form factors in the four-quark scenario
is beyond the conventional quark model.

\subsection{{\it CP} violation}
Thus far \CP violation has not been observed in any $B$ decays involving a scalar meson. The predictions based on QCD factorization are summarized in Tables \ref{tab:asySP} and \ref{tab:asySV}. Mixing-induced \CP asymmetry in $B\to f_0(980)K_S$ decays has been studied in \cite{CCY:SP,ElBennich:2006yi,Dutta:2008xw}.

\section{Conclusion}

We have studied the hadronic charmless $B$ decays to scalar
mesons within the framework of QCD factorization. The main results are:

\begin{itemize}

\item  We have considered two possible scenarios for the scalar
  mesons above 1 GeV, depending on
  whether the light scalars $K_0^*(800),~a_0(980)$ and $f_0(980)$ are
  treated as the lowest lying $q\bar q$ states or four-quark
  particles. We found that the experimental data favor the scenario in which the scalar mesons $a_0(1450)$ and $K_0^*(1430)$ are the lowest lying $q\bar q$ bound states. This in turn implies a preferred four-quark nature for light scalars below 1 GeV.

\item The data of $B\to K_0^*(1430)\eta^{(')}$ and $B\to K_0^*(1430)(\rho,\omega,\phi)$ can be accommodated within the framework of QCD factorization without introducing power corrections from penguin annihilation, while the predicted $B^-\to \ov K_0^{*0}(1430)\pi^-$ and $\ov B^0\to K_0^{*-}(1430)\pi^+$ are too small compared to experiment. In view of the fact that the calculated $K_0^*\rho$ rates are in good agreement with experiment,  it is very important to have more accurate measurements of $B\to K_0^*\pi$ decays to pin down the discrepancy between theory and experiment for $K_0^*\pi$ modes.

\item If $K_0^*(1430)$ is made of the lowest-lying $q\bar q$, we found that Figs. 1(a) and 1(c) interfere constructively and that $A(B\to K_0^*\eta)/A(B\to K_0^*\eta')\approx \cot\!\phi$ with $\phi$ being the $\eta$-$\eta'$ mixing angle in the $\eta_q$ and $\eta_s$ flavor basis. Hence,
    $K_0^*\eta$ has a rate slightly larger than $K_0^*\eta'$ owing to the fact that $\phi$ is less than $45^\circ$. This  is in sharp contrast to the $B\to K\eta'$ decay which has the largest rate in 2-body decays of the $B$ meson.

\item To accommodate the data of $B\to K_0^*(1430)(\eta,\eta')$ and $B\to K_0^*(1430)\phi$ we found that a small form factor for $B\to K_0^*(1430)$ transition as obtained in the covariant light-front quark model is preferable, though other approaches such as pQCD and QCD sum rules tend to yield large form factors for $B$ to $S$ transitions.

\item We have corrected the results for $a_0(980)K$ and $a_0(1450)K$ modes obtained in the previous study \cite{CCY:SP}. Branching fractions should be of order $10^{-7}$ for $B\to a_0(980)K$ in scenario 1 and of order $10^{-6}$ for $B\to a_0(1450)K$ in scenario 2 rather than $10^{-6}$ and $10^{-7}$, respectively, as predicted before. It is expected in QCDF that $a_0(980)^+K^-$ has a rate much smaller than $a_0(980)^+\pi^-$, whereas it is the other way around in pQCD. Experimental measurements of these two modes will help discriminate between these two different approaches.

\item  Although it is widely perceived that light scalar mesons such as
   $f_0(980)$, $a_0(980)$ are predominately four-quark states, in practice it is difficult to make quantitative predictions on $B\to SP$ based on the four-quark picture for $S$. Hence, in practice we shall assume the two-quark scenario for light scalar mesons in calculations.
   So far the calculated  $B\to a_0(980)\pi$ and $B\to a_0(980)K$ rates  in QCDF are all consistent with the experimental limits.  Hence, we cannot conclude on the 2-quark or 4-quark nature of $a_0(980)$.
   Nevertheless,
   if the branching fraction of  $\ov B^0\to a_0^\pm(980)\pi^\mp$ rate is found to be smaller, say, of order $1\times 10^{-6}$, it could imply a four-quark assignment for $a_0(980)$. Since  $\B(\ov B^0\to a_0^\pm(980)\rho^\mp)$ is predicted to be large in QCDF, of order $23\times 10^{-6}$, a measurement of this mode may give the first observation of the $a_0(980)$ production in $B$ decays.

\item  Assuming 2-quark bound states for $f_0(980)$ and $f_0(500)$, the observed large rates of $f_0(980)K$ and $f_0(980)K^*$ modes can be explained in QCDF with the $f_0(980)\!-\!f_0(500)$ mixing angle $\theta$ in the vicinity of $20^\circ$. However, this does not necessarily imply that a 4-quark nature for $f_0(980)$ is ruled out because of extra diagrams contributing to $B\to f_0(980)K^{(*)}$.
    Irrespective of the mixing angle $\theta$, the predicted rate of $B\to f_0(980)\rho^0$ is far below the Belle measurement and this needs to be clarified in the future.

\item Contrary to the odd-parity meson sector, we found that penguin-annihilation induced power corrections are not needed to explain the penguin dominated $B\to SP$ and $B\to SV$ decays in QCDF except for $K_0^*(1430)\pi$ modes. How to understand both $K_0^*(1430)\pi$ and $K_0^*(1430)\rho$ simultaneously remains an issue in QCD-inspired approaches.

\end{itemize}

\vskip 2.5cm \acknowledgments

This research was supported in part by the National Science
Council of R.O.C. under Grant Nos. NSC99-2112-M-033-005-MY3, NSC100-2112-M-001-009-MY3 and NSC100-2112-M-033-001-MY3 and by the National Natural Science Foundation of China under Grant
No. 11147004.

\appendix

\section{Decay amplitudes of $B\to SP,SV$}
Within the framework of QCD factorization, the decay amplitudes of $B\to SP, SV$ decays can be found in \cite{CCY:SP} and \cite{CCY:SV}, respectively. However, some factorizable terms were missed in the expressions for the
decay amplitudes of $B\to (f_0,a_0^0)(K,\pi)$ given before in \cite{CCY:SP}; that is,
we did not take into account the contributions from
the $f_0$ or the neutral $a_0$ emission induced from the
four-quark operators other than $O_6$ and $O_8$ (see also
\cite{Shen}). They are corrected here: \footnote{
In Eq. (\ref{eq:SDAmp}), when applying bar to $\alpha_4$ and $\alpha_{\rm 4, EW}$, we do not apply bar to $a_6$ and $a_8$; that is, for those Wilson coefficients associated with $r_\chi$, the bar applies only to $r_\chi$ itself.
For example, $\bar \alpha_4$=$\bar a_4\mp \bar r_\chi a_6$.
The $\bar r_\chi a_{6,8}$ terms in Eq. (\ref{eq:SDAmp}) are in agreement with those in the formulas in  Appendix A of \cite{CCY:SP}.}

\be \label{eq:SDAmp}
A(B^- \to f_0 K^- ) &=&
 \frac{G_F}{\sqrt{2}}\sum_{p=u,c}\lambda_p^{(s)}
 \Bigg\{ \left(\bar \alpha_3^p+\bar\alpha_4^p-{1\over 2}\bar\alpha_{\rm 3,EW}^p-{1\over 2}\bar\alpha_{\rm 4,EW}^p\right)_{Kf_0^s}\ov X^{(\bar BK,f_0^s)}
 \non \\
 &+& \left(a_1 \delta^p_u+\alpha_4^p
 +\alpha_{\rm 4,EW}^p \right)_{f_0^u K} X^{(\bar B f_0^u, K)} + \left(\bar a_2\delta^p_u+2\bar\alpha_3^p+{1\over 2}\bar\alpha_{\rm 3,EW} \right)_{Kf_0^u} \ov X^{(\bar BK,f_0^u)}
 \non \\
 &+& f_Bf_K\bigg[ \bar f_{f_0^u}\big(\bar b_2\delta_u^p+\bar b_3
 +\bar b_{\rm 3,EW}\big)_{f_0^uK} +\bar f_{f_0^s}\big(\bar b_2\delta_u^p+\bar b_3
 +\bar b_{\rm 3,EW}\big)_{Kf_0^s}\bigg] \Bigg\}, \non \\
A(\ov B^0 \to f_0\ov K^0 ) &=&
 ¥ú\frac{G_F}{\sqrt{2}}\sum_{p=u,c}\lambda_p^{(s)}
 \Bigg\{ \left(\alpha_3^p+\alpha_4^p-{1\over 2}\alpha_{\rm 3,EW}^p-{1\over 2}\alpha_{\rm 4,EW}^p\right)_{Kf_0^s}\ov X^{(\bar BK,f_0^s)}
  \non \\
 &+& \left(\bar\alpha_4^p
 -{1\over 2}\bar\alpha_{\rm 4,EW}^p\right)_{f_0^dK}X^{(\bar B f_0^u, K)}+\left(\bar a_2\delta^p_u+2\bar\alpha_3^p+{1\over 2}\bar\alpha_{\rm 3,EW} \right)_{Kf_0^u} \ov X^{(\bar BK,f_0^u)}
 \non \\
  &+& f_Bf_K\bigg[\bar f_{f_0^d}\big(\bar b_3
 -{1\over 2}\bar b_{\rm 3,EW}\big)_{f_0^dK} + \bar f_{f_0^s}\big(\bar b_3
 -{1\over 2}\bar b_{\rm 3,EW}\big)_{Kf_0^s}\bigg] \Bigg\}, \non \\
A(B^- \to a_0^0 K^- ) &=&
\frac{G_F}{{2}}\sum_{p=u,c}\lambda_p^{(s)}
 \Bigg\{ \left(a_1 \delta^p_u+\alpha_4^p
 +\alpha_{\rm 4,EW}^p \right)_{a_0 K} X^{(\bar Ba_0,K)} \non \\
 &+& (\bar a_2)_{K a_0} \ov X^{(\bar BK,a_0)}
 + f_Bf_K\bar f_{a_0}\big(\bar b_2\delta_u^p+\bar b_3
 +\bar b_{\rm 3,EW}\big)_{a_0K}\Bigg\}, \non \\
A(\ov B^0 \to a_0^0\ov K^0 ) &=&
 \frac{G_F}{{2}}\sum_{p=u,c}\lambda_p^{(s)}
 \Bigg\{ \left(\alpha_4^p
 -{1\over 2}\alpha_{\rm 4,EW}^p \right)_{a_0K}X^{(\bar Ba_0,K)} \non \\
 &+& (\bar a_2)_{K a_0} \ov X^{(\bar BK,a_0)}
  + f_Bf_K \bar f_{a_0}\big(\bar b_3
 -{1\over 2}\bar b_{\rm 3,EW}\big)_{a_0K} \Bigg\}, \non \\
A(B^- \to f_0 \pi^- ) &=&
 \frac{G_F}{\sqrt{2}}\sum_{p=u,c}\lambda_p^{(s)}
 \Bigg\{ \left(\bar a_1 \delta^p_u+\bar\alpha_4^p
 +\bar\alpha_{\rm 4,EW}^p \right)_{f_0^u \pi} X^{(\bar Bf_0^u,\pi)} \non \\ &+&
 \left(\bar a_2 \delta^p+2\bar\alpha_3^p+{1\over 2}\bar\alpha_{\rm 3,EW}^p+\bar\alpha_4^p-{1\over 2}\bar\alpha_{\rm 4,EW}^p\right)_{\pi f_0^u}\ov X^{(\bar B\pi,f_0^u)}\non \\
 &-& f_Bf_\pi\bar f_{f_0^u}\bigg[\big(\bar b_2\delta_u^p+\bar b_3
 +\bar b_{\rm 3,EW}\big)_{f_0^u \pi} +\big(\bar b_2\delta_u^p+\bar b_3
 +\bar b_{\rm 3,EW}\big)_{\pi f_0^u}\bigg] \Bigg\}, \non \\
A(\ov B^0 \to f_0\pi^0 ) &=&
\frac{G_F}{2}\sum_{p=u,c}\lambda_p^{(s)}
 \Bigg\{ \left(-a_2 \delta^p_u+\alpha_4^p-{3\over 2}\alpha_{\rm 3,EW} -{1\over 2}\alpha_{\rm 4,EW} \right)_{f_0^d \pi} X^{(\bar Bf_0^u,\pi)} \non \\
 &+&
  \left(\bar a_2\delta^p_u+2\bar\alpha_3^p+\bar\alpha_4^p+{1\over 2}\bar\alpha_{\rm 3,EW}^p-{1\over 2}\bar\alpha_{\rm 4,EW}^p\right)_{\pi f_0^d}\ov X^{(\bar B\pi,f_0^d)} - f_Bf_\pi\bar f_{f_0^d}\non \\
  &\times& \bigg[\big(\bar b_1 \delta^p_u-\bar b_3
 +{1\over 2}\bar b_{\rm 3,EW}+{3\over 2}\bar b_{\rm 4,EW}\big)_{f_0^d \pi} +\big(\bar b_1 \delta^p_u-\bar b_3
 +{1\over 2}\bar b_{\rm 3,EW}+{3\over 2}\bar b_{\rm 4,EW}\big)_{\pi f_0^d}\bigg] \Bigg\},\non \\
  A(B^- \to a^0_0\pi^- ) &=&
\frac{G_F}{2}\sum_{p=u,c}\lambda_p^{(d)}
 \Bigg\{ \left( a_1\delta_u^p+ \alpha_4^p
 +\alpha_{\rm 4,EW}^p \right)_{a_0\pi} X^{(\bar Ba_0,\pi)}\non \\
 &+&
 \left(\bar a_2\delta^p_u-\bar\alpha_4^p+{1\over 2}\bar\alpha_{\rm 4,EW}^p+{3\over 2}\bar\alpha_{\rm 3,EW}^p \right)_{\pi a_0} \ov X^{(\bar B\pi,a_0)} \non \\
 &+& f_Bf_\pi\bar f_{a_0}\Big[\big(\bar b_2\delta_\mu^p+\bar b_3+
 \bar b_{\rm 3,EW}\big)_{a_0\pi}-
 \big(\bar b_2\delta_\mu^p+\bar b_3 +\bar b_{\rm 3,EW}\big)_{\pi a_0} \Big]
 \Bigg\}, \non \\
  A(\ov B^0 \to a^0_0\pi^0 ) &=&
-\frac{G_F}{2\sqrt{2}}\sum_{p=u,c}\lambda_p^{(d)}
 \Bigg\{ \left( a_2\delta_u^p- \alpha_4^p
 +{1\over 2}\alpha_{\rm 4,EW}^p+{3\over 2}\alpha_{\rm 3,EW}^p \right)_{a_0\pi} X^{(\bar Ba_0,\pi)} \non \\
 &+&
 \left(\bar a_2\delta^p_u -\bar\alpha_4+{1\over 2}\bar \alpha_{\rm 4,EW}^p+{3\over 2}\bar\alpha_3^p\right)_{\pi a_0}\ov X^{(\bar B\pi,a_0)} \non \\
 &-& f_Bf_\pi\bar f_{a_0}\Big[\big(\bar b_1\delta_\mu^p+\bar b_3+2\bar b_4-
 {1\over 2}(\bar b_{\rm 3,EW}-\bar b_{\rm 4,EW})\big)_{a_0\pi} \non \\
 &+& \big(\bar b_1\delta_\mu^p+\bar b_3+2\bar b_4 - {1\over 2}(\bar b_{\rm 3,EW}-\bar b_{\rm 4,EW})
 \big)_{\pi a_0} \Big]
 \Bigg\}.
 \en
Here $a_0$ stands for $a_0(980)$ or $a_0(1450)$, $\lambda_p^{(q)}\equiv V_{pb}V_{pq}^*$ with $q=d,s$,
 \be \label{eq:r}
 && r^K_\chi(\mu)={2m_K^2\over m_b(\mu)(m_u(\mu)+m_s(\mu))}, \qquad
 r^{K^*_0}_\chi(\mu)={2m_{K_0^*}^2\over
 m_b(\mu)(m_s(\mu)-m_q(\mu))}, \non \\
 &&  r^{a_0}_\chi(\mu)={2m_{a_0}^2\over
 m_b(\mu)(m_d(\mu)-m_u(\mu))}, \quad \bar r_\chi^{a_0}(\mu)={2m_{a_0}\over
 m_b(\mu)},  \quad \bar r_\chi^{f_0}(\mu)={2m_{f_0}\over m_b(\mu)},
 \en
and
\be \label{eq:replacementII}
 \alpha_3^p(M_1M_2) &=& \cases{ a_3^p(M_1M_2)+a_5^p(M_1M_2); & for ~$M_1M_2=PS,SV,VS$, \cr
 a_3^p(M_1M_2)-a_5^p(M_1M_2); & for ~$M_1M_2=SP$,} \non \\
 \alpha_4^p(M_1M_2) &=& \cases{ a_4^p(M_1M_2)-r_\chi^{M_2} a_6^p(M_1M_2); & for ~$M_1M_2=PS,SP,SV$, \cr
 a_4^p(M_1M_2)+r_\chi^{M_2} a_6^p(M_1M_2); & for ~$M_1M_2=VS$,}  \\
 \alpha_{\rm 3,EW}^p(M_1M_2) &=& \cases{ a_9^p(M_1M_2)+a_7^p(M_1M_2); & for ~$M_1M_2=PS,SV,VS$, \cr
 a_9^p(M_1M_2)-a_7^p(M_1M_2); & for ~$M_1M_2=SP$,} \non \\
  \alpha_{\rm 4,EW}^p(M_1M_2) &=& \cases{ a_{10}^p(M_1M_2)-r_\chi^{M_2} a_8^p(M_1M_2); & for ~$M_1M_2=PS,SP,SV$, \cr
 a_{10}^p(M_1M_2)+r_\chi^{M_2} a_8^p(M_1M_2); & for ~$M_1M_2=VS$. }
 \non
\en

The general expression of the effective Wilson coefficients $a_i$ is given in Eq. (\ref{eq:ai}), while the effective Wilson coefficients $\bar a_i$ appearing in Eq. (\ref{eq:SDAmp}) are defined as $a_i \mu^{-1}_S$  and they can be obtained from Eq.
(\ref{eq:ai}) by retaining only those terms that are proportional
to $\mu_S$. Specifically,
 \be \label{eq:barai}
 \bar a_i^p(M_1M_2) ={c_{i\pm1}\over N_c}\,{C_F\alpha_s\over
 4\pi}\Big[\bar V_i(M_2)+{4\pi^2\over N_c}\bar H_i(M_1M_2)\Big]+\bar
 P_i^p(M_2).
 \en
The LCDA of the neutral scalar meson  is replaced by
$\bar\Phi_S$ which has the similar expression as Eq.
(\ref{eq:twist2wf}) except that the first constant term does not
contribute and the term $f_S\,\mu_S$ is factored out:
 \be
 \bar \Phi_S(x,\mu)= 6x(1-x)\sum_{m=1}^\infty
 B_m(\mu)\,C_m^{3/2}(2x-1).
 \en
The annihilation terms $\bar b_i$ have the same expressions as $b_i$ except that $r_\chi^S$ and $\mu_S B_i$ are replaced by $\bar
r_\chi^S$ and $B_i$, respectively (see \cite{CCY:SV}).

In Eq. (\ref{eq:SDAmp}), the quantity $X$ is a factorizable term whose explicit expression is given by
\be
\label{eq:X}
X^{(\bar BS,V)} &\equiv & \la V| J^{\mu}|0\ra\la
S|J'_{\mu}|\ov B \ra=2f_V\,m_Bp_c F_1^{ B
S}(m_{V}^2),   \non \\
X^{( \bar BV,S)} &\equiv &
\la S | J^{\mu}|0\ra\la V|J'_{\mu}|\ov B
\ra=-2f_S\,m_Bp_cA_0^{B V}(m_{S}^2),   \\
X^{( \bar BS,P)} &\equiv &
\la P | J^{\mu}|0\ra\la S|J'_{\mu}|\ov B
\ra=-f_P\,(m_B^2-m_S^2)F_0^{B S}(m_{P}^2), \non \\
X^{(\bar BP,S)} &\equiv & \la S| J^{\mu}|0\ra\la
P|J'_{\mu}|\ov B \ra=f_S\,(m_B^2-m_P^2)F_0^{ B
P}(m_{S}^2), \non
\en
where $p_c$ is the c.m. momentum.

Note that the $f_0(980)$--$f_0(500)$ mixing angle (i.e. $\sin\theta$) and
Clebsch-Gordon coefficient $1/\sqrt2$ have been included in the
$f_0(980)$ form factors $F^{Bf^{u,d}_0}$ and decay constants
$f_{f_0}^{u,d}$.
Throughout, the order of the arguments of the $a_i^p(M_1M_2)$ and
$b_i(M_1M_2)$ coefficients is dictated by the subscript $M_1M_2$,
where $M_2$ is the emitted meson and $M_1$ shares the same
spectator quark with the $B$ meson. For the annihilation diagram,
$M_1$ is referred to the one containing an antiquark from the weak
vertex, while $M_2$ contains a quark from the weak vertex.


\end{document}